\documentclass[11pt,letter]{article}
\usepackage[english]{babel}
\usepackage[utf8x]{inputenc}
\usepackage[T1]{fontenc}
\usepackage{preamble}

\title{Pure Differential Privacy from Secure Intermediaries}
\author{Albert Cheu \& Chao Yan\\ Department of Computer Science\\ Georgetown University}

\begin{document}

\maketitle
\abstract{Recent work in differential privacy has explored the prospect of combining local randomization with a secure intermediary. Specifically, there are a variety of protocols in the \emph{secure shuffle} model (where an intermediary randomly permutes messages) as well as the \emph{secure aggregation} model (where an intermediary adds messages). Most of these protocols are limited to approximate differential privacy. An exception is the shuffle protocol by Ghazi, Golowich, Kumar, Manurangsi, Pagh, and Velingker: it computes bounded sums under pure differential privacy. Its additive error is $\tilde{O}(1/\eps^{3/2})$, where $\eps$ is the privacy parameter. In this work, we give a new protocol that ensures $O(1/\eps)$ error under pure differential privacy. We also show how to use it to test uniformity of distributions over $[d]$. The tester's sample complexity has an optimal dependence on $d$. Our work relies on a novel class of secure intermediaries which are of independent interest.}

\section{Introduction}
Following the seminal work by Dwork, McSherry, Nissim, and Smith \cite{DMNS06}, differential privacy research has yielded much fruit in the central model. There, the party who wishes to compute on data---the \emph{analyzer}---has direct access to the data. This is not desirable when the party is not trusted to enforce privacy. The local model addresses this limitation: each owner of data---each \emph{user}---takes it upon themselves to ensure privacy via a \emph{local randomizer}. Because each user injects a non-trivial amount of randomness, computations are noisier in the local model than the central model. Consider the bounded sum problem: each user has a value in the interval $[0,1]$ and the analyzer wants an estimate of their sum. When $\eps$ is the target privacy parameter and $n$ is the number of users, $\Omega(\sqrt{n}/\eps)$ error is necessary in the local model but $O(1/\eps)$ error is possible in the central model \cite{CSS12,BNO08,DMNS06}.

In response to the limitations of the local model, there is a growing body of work on alternative distributed models. In particular, the secure shuffle model and the secure aggregation model have gathered considerable interest (see e.g. \cite{GXS13, SCRCS11, KLS21, Cheu21}). Like in the local model, users perform some randomization on their data but these models further assume the existence of an intermediary that computes some (possibly randomized) function $f$ on user messages and reports the output to the analyzer. For example, the shuffler is the functionality that applies a uniformly random permutation to $\vec{y}$. Meanwhile, the secure aggregation literature studies implementations of modular addition.


Some protocols using these (ideal-world) intermediaries outperform any counterpart in the local model. For example, Balle, Bell, Gasc\'{o}n, and Nissim \cite{BBGN19-2} give a shuffle protocol with expected error $O(1/\eps)$, matching the bound in the central model.

One limitation of that protocol is that the additive privacy parameter $\delta$ must be strictly positive. This is known as \emph{approximate differential privacy}. Other shuffle protocols for bounded sums and binary sums also demand $\delta>0$ \cite{CSU+19, GGK+19, BC20, BCJM21}. Meanwhile, the Laplace mechanism in the central model produces estimates with $O(1/\eps)$ error while satisfying \emph{pure differential privacy}, where $\delta=0$.

Ghazi, Golowich, Kumar, Manurangsi, Pagh, and Velingker \cite{GGK+20} do give a shuffle protocol for bounded sums that satisfies pure differential privacy, but they are only able to bound the error by $\tilde{O}(1/\eps^{3/2})$. This prompts the following question:

\begin{quote}
Is there a shuffle protocol for bounded sums that satisfies $\eps$-differential privacy and has expected error $O(1/\eps)$?
\end{quote}

\paragraph{Our Contributions.} 
We answer this question in the affirmative. In the first phase of our construction, we describe a local randomizer such that shuffling the outputs from $n$ executions instantiates an aggregator with \emph{relaxed} properties. Specifically, it miscalculates the modular sum with a bounded probability and ensures an adversary can only learn $\eps$ more than that sum. Rather than bounding the statistical distance between the adversary's view and an ideal-world simulation, $\eps$ bounds a metric derived from likelihood ratios.

In the second phase, we create a protocol for bounded sums via the relaxed aggregator. Roughly speaking, we describe a randomizer such that the modular sum of $n$ executions is $\eps$-private and has $O(1/\eps)$ error (it will add noise drawn from the discrete Laplace distribution). Because the relaxed aggregator's security property is in terms of likelihood ratios, it is compatible with pure differential privacy: the composition of randomizer and aggregator is $O(\eps)$-private.

The protocol for bounded sums serves as a building block for testing uniformity of probability distributions. Focusing on the influence of dimensionality $d$, the sample complexity of this tester is $O(d^{2/3})$. This matches an existing lower bound.

We remark that our relaxed aggregator is one instance of a general class of intermediary which could be of independent interest. Because we have replaced statistical distance with an alternate metric, there is potential for theory and practice to engineer protocols and primitives with novel guarantees.

\subsection{Prior Work}
Below, we place our results in context with prior work.

\begin{table}[h]
    \centering
    \begin{tabular}{cccc}
         Problem & Source & Bound in the Shuffle Model & Privacy Type \\ \hline
         & & Additive Error & \\ 
            Bounded & \cite{BBGN19-2} & $O(1/\eps )$ & Approximate \\
         Sums & \cite{GGK+20} & $O(\sqrt{\log (1/\eps)}/\eps^{3/2} )$ & Pure \\
         & Thm. \ref{thm:bounded-sum} & $O(1/\eps )$ & Pure \\ \hline
         & & Sample Complexity & \\ 
         $\alpha$-Uniformity & \cite{CL21} & $O\paren{ \frac{d^{2/3}}{\alpha^{4/3}\eps^{2/3}} \log^{1/3}\frac{1}{\delta} + \frac{\sqrt{d}}{\alpha\eps} + \frac{\sqrt{d}}{\alpha^2} }$ & Approximate \\
         Testing & Thm. \ref{thm:ut-final} & $O\paren{\frac{d^{2/3}}{\alpha^{4/3}\eps^{2/3}} + \frac{\sqrt{d}}{\alpha\eps} + \frac{\sqrt{d}}{\alpha^2}}$ & Pure \\
         & \cite{BCJM21} & $\Omega\paren{\frac{d^{2/3}}{\alpha^{4/3}\eps^{2/3}} + \frac{1}{\alpha\eps} + \frac{\sqrt{d}}{\alpha^2}}$ & Pure \\ \hline
    \end{tabular}
    \captionsetup{width=0.85\textwidth,font=small}
    \caption{Results for bounded sums and uniformity testing in the shuffle model. $\alpha$ is an accuracy parameter and $d$ is the size of the domain; refer to Definition \ref{defn:ut} for specifics.}
    \label{tab:sums}
\end{table}

Shuffle protocols for differential privacy can be traced back to work by Bittau et al. \cite{BittauEMMRLRKTS17} and Cheu et al. \cite{CSU+19}. Protocols for bounded-value sums can be found in the latter work, as well as the follow up works by Balle, Bell, Gasc{\'{o}}n, and Nissim \cite{BBGN19} \& Ghazi, Pagh, Velingker \cite{GPV19}. As previously stated, Ghazi et al. \cite{GGK+20} give a shuffle protocol for the problem that satisfies pure differential privacy, but with an undesirable dependence on $\eps$.

Differential privacy via secure aggregation constitute another line of research. The seeds can be found in early work by Dwork Kenthapadi, McSherry, Mironov, and Naor \cite{DKMMN06}. Recent work by Kairouz, Liu, and Steinke \cite{KLS21} (resp. Agarwal, Kairouz, and Liu \cite{AKL21}) give a communication-efficient protocol for vector sums that relies on the discrete Gaussian distribution (resp. Skellam distribution). Interestingly, the work by Bell, Bonawitz, Gasc\'{o}n, Lepoint, and Raykova \cite{BBGLR20} not only give a cryptographic instantiation of a secure aggregator but also show how to simulate a shuffler with one. Our work provides a construction that essentially performs the reverse simulation.

Our approach to compute bounded sums in the shuffle model most resembles the one by Balle et al. \cite{BBGN19-2}. We summarize the main ideas below:
\begin{enumerate}
    \item Construct an aggregator in the shuffle model that is perfectly correct but statistically secure. That is, for any two input vectors $\vec{x},\vec{x}\,'$ such that $\sum x_i = \sum x'_i$ modulo $m$, an adversary's view only slightly differs when changing between $\vec{x}$ and $\vec{x}\,'$. The difference is quantified in statistical distance.
    \item Use that aggregator to build an $(\eps,\delta)$-differentially private protocol for summation of bounded values. Each user adds a small amount of independent noise to their data and then sends the noised value to the aggregator. The view of an adversary is insensitive to a user's contribution due to the (discrete Laplace) noise accumulated from all users.
\end{enumerate}
Our protocol primarily differs in the first step: our aggregator's security guarantee is defined in terms of a metric based on likelihood ratios rather than statistical distance.

This alternative metric was also studied in prior work by Haitner,  Mazor, Shaltiel, and Silbak \cite{HMSS19}. There, the authors construct oblivious transfer from an agreement primitive. In our work, we construct counting protocols from shuffling and aggregation primitives.

Our definition of robust differential privacy is essentially the same as the one in work by \'{A}cs and Castelluccia \cite{AC11}. Balcer, Cheu, Joseph, and Mao \cite{BCJM21} give a more general definition which allows privacy parameters to depend on the number of corruptions.

Uniformity testing under pure differential privacy is well-studied. Acharya, Canonne, Freitag, and Tyagi \cite{ACFT19} show that the sample complexity in the local model is necessarily linear in the dimension $d$. In contrast, a folklore upper bound in the central model is $O(\sqrt{d}/\alpha^2\eps)$; a sketch can be found in Section 3 of work by Cai, Daskalakis, and Kamath \cite{CDK17}. That work also contains an upper bound with improved dependence on the privacy and accuracy parameters, later refined by Aliakbarpour, Diakonikolas, and Rubinfeld \cite{ADR18} and Acharya, Sun, and Zhang \cite{ASZ18}. For the more exotic pan-private model, Amin, Joseph, and Mao \cite{AJM20} give upper and lower bounds that scale with $d^{2/3}$. Balcer et al. \cite{BCJM21} derive a lower bound for shuffle privacy by transforming a robustly private shuffle protocol into a pan-private one, so that the lower bound from Amin et al. \cite{AJM20} carries over.

Balcer et al. also give a robustly private shuffle protocol for testing that demands $\tilde{O}(d^{2/3})$ samples. Their arguments are streamlined by follow-up work by Canonne and Lyu \cite{CL21}, who also describe an alternative protocol that consumes only one message. However, both of these protocols only satisfy approximate differential privacy. Our shuffle protocol satisfies pure differential privacy and its sample complexity scales with $d^{2/3}$, which matches the lower bound by Balcer et al.

\section{Preliminaries}

For probability distribution $\bD$ and event $E$, we use ``$\pr{}{\bD\in E}$'' as shorthand for $\pr{\eta \sim \bD}{\eta \in E}$. For any pair of distributions $\bD,\bD'$, let $\supp(\bD,\bD')$ be the union of their supports. The statistical distance is $$\sd{\bD}{\bD'}:= \max_{E\in \supp(\bD,\bD')} |\pr{}{\bD \in E}- \pr{}{\bD' \in E}|.$$ But much of our work will center on the following metric:
\begin{defn}
The \emph{log-likelihood-ratio (LLR) distance} between $\bD,\bD'$ is $$ \llr(\bD, \bD') := \max_{E\in\supp(\bD,\bD')} \left| \ln \left(  \frac{ \pr{}{\bD \in E }}{\pr{}{\bD' \in E}} \right) \right|,$$
where division by zero (resp. $\ln(0)$) is treated as $+\infty$ (resp. $-\infty$).
\end{defn}

Differential privacy can be defined in terms of this metric:
\begin{defn}
An algorithm $\cM$ is $\eps$-\emph{differentially private} if, for any pair of inputs $\vec{x},\vec{x}\,' \in \cX^n$ that differ on one user, $$\llr(\cM(\vec{x}),\cM(\vec{x}\,')) \leq \eps.$$ We also say that $\cM$ satisfies \emph{pure differential privacy}.
\end{defn}
Throughout this work, we will assume we are in the regime $\eps=O(1)$ so that $e^\eps-1$ (resp. $e^\eps$) can be written as $O(\eps)$ (resp. $O(1)$).

One way to ensure central differential privacy is with the use of discrete Laplace noise (also known as symmetric geometric noise). Let $\DLap(\mu,\rho)$ be the distribution with mean $\mu$ and scale parameter $\rho \in (0,1)$. The mass it places on any integer $v$ is proportional to $\rho^{|v-\mu|}$, written as $\pr{}{\DLap(\mu,\rho)=v}\propto \rho^{|v-\mu|}$. In the case where $\mu=0$, we simply write $\DLap(\rho)$.

\begin{thm}[Discrete Laplace Mechanism]
\label{thm:lap-mech}
Let $f:\cX^n\to \Z$ be a $\Delta$-sensitive function. Let $M_f:\cX^n\to \Z$ be the algorithm that, on input $\vec{x}$, samples $\eta\sim \DLap(e^{-\eps / \Delta})$ and reports $f(\vec{x}) + \eta$. $M_f$ is $\eps$-differentially private and reports values with a standard deviation of error $O(\Delta/\eps)$.
\end{thm}

A useful property of the discrete Laplace distribution is \emph{infinite divisibility}:
\begin{fact}
\label{fact:polya-dlap}
For any $n\in\N$, let $\{\eta^{+}_i,\eta^{-}_i\}_{i \in [n]}$ be $2n$ independent samples from the distribution $\Polya(1/n,\rho)$. The random variable $\sum_{i=1}^n \eta^{+}_i - \eta^{-}_i$ is distributed as $\DLap(\rho)$.
\end{fact}

For proofs of Theorem \ref{thm:lap-mech} and Fact \ref{fact:polya-dlap}, we refer the interested reader to the prior work by Balle, Bell, Gasc\'{o}n, and Nissim \cite{BBGN19-2} and the citations within.

Our work will also use the \emph{truncated} discrete Laplace distribution, denoted $\DLap_\tau(\mu,\rho)$. If $v\notin [\mu-\tau,\mu+\tau]$ then it places zero mass on $v$ and otherwise the mass is $\propto \rho^{|v-\mu|}$. As before, $\DLap_\tau(\rho)$ will be shorthand for $\DLap_\tau(0,\rho)$.

\subsection{The Secure Intermediary Model}

Here, we give notation for a model which abstracts the shuffle and aggregation models from prior work. $n$ parties $i\in[n]$ called \emph{users} own data $x_1,\dots,x_n$. One party called the \emph{analyzer} wishes to compute on that data. Finally, there is an entity or service called the \emph{secure intermediary} $I$. It runs some algorithm on the messages sent by users and forwards the result to the analyzer; in a slight abuse of notation, $I$ will also refer to the algorithm run by the intermediary.

A protocol $P$ in the \emph{secure intermediary model} consists of a triple $(\vec{R},I,A)$, where $R_1,\dots,R_n,A$ are randomized algorithms and $I$ is a secure intermediary. In the case where $R_1=\dots=R_n$, the protocol is \emph{symmetric} and we simply write $P=(R,I,A)$. Meanwhile, an attack against $P$ is identified by a triple $(C, W)$, where $C \subset [n]$ is a set of corrupted users and $W=\{w_i\}_{i\in C}$ are the messages they send (or $\bot$ if they drop out). We use $\vec{y}_C$ (resp. $\vec{y}_{\overline{C}}$) to denote those values in $\vec{y}$ whose indices lie in $C$ (resp. $\overline{C}$).

\begin{defn}
An execution of $P=(\vec{R},I,A)$ under attack $T=(C,W)$ proceeds as follows:
\begin{enumerate}
    \item If $i \in C$ then user $i$ computes $y_i \gets w_i$ and otherwise $y_i\gets R_i(x_i)$.
    \item User then $i$ sends $y_i$ to $I$ (if it is not $\bot$).
    \item $I$ computes on the vector of messages $\vec{y}$ and sends its output to the analyzer.
    \item The analyzer executes $A$ on the value it receives from $I$.
\end{enumerate}
We define the output of the protocol as the output of the analyzer. Absent an attack, it is written as $P(\vec{x}) := (A\circ I\circ \vec{R})(\vec{x})$. In the case of a symmetric protocol, we write $P(\vec{x}) = (A\circ I\circ R^n)(\vec{x})$.

The adversary only has access to the users under its control and the output of the intermediary, so its view is $\view_T(P,\vec{x}) := (C,W,\vec{x}_C, I(\vec{y}))$.
\end{defn}

We will design protocols in this model that will have robust privacy guarantees: the adversary's view will be insensitive to any small change in the contents of the input vector, even when (a minority of) users launch an attack. We formally define our objective below:

\begin{defn}
\label{defn:robust-dp}
A secure intermediary protocol $P=(\vec{R},I,A)$ is $\eps$-\emph{robustly differentially private} if the following inequality holds for any $T=(C,W)$ where $|C|\leq n/2$:
$$
\llr( \view_T(P,\vec{x}) , \view_T(P,\vec{x}\,') ) \leq \eps.
$$
where $\vec{x},\vec{x}\,' \in \cX^n$ are any pair of inputs that differ on one index $i\notin C$.
\end{defn}

\subsection{Correctness and Security of Intermediaries}
Ideally, $I$ computes some functionality $f$ exactly. But we will see that pure differential privacy remains feasible when we relax this constraint. The relaxed version should preserve \emph{correctness}: on any input $\vec{y}$, an analyzer should be able to obtain an approximate version of $f(\vec{y})$.
\begin{defn}
An intermediary $I$ is $q$-\emph{correct with respect to $f$} if there exists some post-processing function $\POST$ such that
$$
\SD(\POST(I(\vec{y})), f(\vec{y})) \leq q
$$
\end{defn}

We would also like the intermediary to be \emph{secure}: the analyzer should not learn much more about $\vec{y}$ than $f(\vec{y})$.



\begin{defn}
An intermediary $I$ is $\eps$-\emph{log-likelihood-ratio (LLR) secure with respect to} $f:\cY^* \to \cR$ if there exists a simulator $\SIM$ such that, for any input $\vec{y} \in \cY^*$,
$$
\llr(I(\vec{y}), \SIM(f(\vec{y}))) \leq \eps.
$$
\end{defn}

We remark that alternate definitions include perfect security and statistical security. The former ensures that the adversary's view can be \emph{exactly} simulated from $f(\vec{y})$, while the latter permits a small amount of leakage measured in statistical distance. We choose $\llr$ distance in order to be compatible with pure differential privacy.

We package correctness and security into one objective.
\begin{defn}
An intermediary $I$ is an $(\eps,q)$-\emph{relaxation of} $f$ if it is both $\eps$-LLR secure an $q$-correct with respect to $f$.
\end{defn}

In the case where $\eps=0$ and $q=0$, $I$ reports a value which is interchangeable with $f(\vec{y})$ from the point of view of the analyzer. For example, it could apply a lossless compression scheme known to the analyzer or re-scale numerical values by a value known to the analyzer.

\subsection{Two Desirable Functionalities}

Let $\Sigma^{d,m,n}$ be the function that takes in at most $n$ row vectors---each of dimension $d$ and containing integers $\in [0,m-1]$---then reports their sum, modulo $m$ in each coordinate. We refer to this function as ``the aggregator.''

Let $S^{d,m,n}$ be the function that has the same input types as aggregator $\Sigma^{d,m,n}$ but, instead of adding them, it concatenates its inputs and uniformly permutes the resulting vector. We refer to this function as ``the shuffler.''

We will sometimes drop the superscripts when the context makes their content clear.

\section{Creating a Relaxed Aggregator from Relaxed Shuffling}
\label{sec:sim-mod}

In this section, we show how to make an $(\eps,q)$-relaxed aggregator by composing local randomization with an $(<\eps, <q$)-relaxed shuffler. This will be the foundation of a protocol for bounded sums (Section \ref{sec:bounded-sum}). We also prove we cannot create an $(0,q)$-relaxed aggregator from the (non-relaxed) shuffler without an inordinately large $q$.

\subsection{Setup and Intuition}
We first present the construction for the one-dimensional case ($\Sigma^{d=1,m,n}$), then explain how to extend it to larger dimensions ($\Sigma^{d>1,m,n}$). The protocol will take parameters $\lambda,p \in (0,1)$ and $t\in \N$. Their values will determine the security $\eps$ and failure rate $q$ of the construction. We will use $\bD$ as shorthand for $\DLap_{t/2}(t/2,\exp(-\lambda))$. It is the distribution over $\{0,\dots,t\}$ such that $\pr{}{\bD =u} \propto \exp(-\lambda \cdot |\tfrac{t}{2} - u|)$. Also for brevity, let $\bD[u] := \pr{}{\bD=u}$.

For $x\in \{0,1,\dots,m-1\}$, let $\bD_x$ be the distribution such that $$\pr{\eta \sim \bD_x}{\eta=x+um} := \bD[u].$$ This is supported on values that have modular residue $x$---e.g. $\bD_0$ has support $0,m,2m,\dots$ while $\bD_1$ has support $1,m+1,2m+1,\dots$---where the mass on a point is determined by the truncated discrete Laplace distribution. We will continue the use of square brackets as shorthand for the probability mass function: we write $\bD_x[x+um] := \bD[u]$.

\paragraph{A First Attempt.} Consider the following randomizer:
\begin{center}
$\overline{R}(x) :=$ binary vector with length $mt+(m-1)$ \& sum $\eta$, where $\eta \sim \bD_x$
\end{center}
Let $\overline{R}^{n}(\vec{x})$ be $n$ independent executions of $\overline{R}$ on each entry of $\vec{x}$. Suppose that we have access to the shuffler $S^{mt+(m-1),2,n}$, which can permute all the bits generated by $\overline{R}^{n}(\vec{x})$. For brevity, we will drop the superscript. Observe that the composition $(S\circ \overline{R}^{n})(\vec{x})$ produces a sequence of ones and zeroes such that their sum has the same modular residue as $\sum x_i$. Thus, we achieve perfect correctness with this randomizer.

Things are less rosy when we turn to security. In the case where $n=2$, consider the pair of inputs $\vec{x}=(0,0)$ and $\vec{x}\,'=(1,m-1)$. Because these have the same sum modulo $m$, $(S\circ \overline{R}^2)(\vec{x})$ should be $2\eps$ close to $(S\circ \overline{R}^2)(\vec{x}\,')$. But there is nonzero probability that $(S\circ \overline{R}^2)(\vec{x})$ produces output that is all zeroes, while the corresponding probability for $(S\circ \overline{R}^2)(\vec{x}\,')$ is zero (since $\overline{R}(1), \overline{R}(m-1)$ place no mass on all-zero vectors). The immediate corollary is that there is no finite $\eps$ for which this construction provides $\eps$-LLR security.

\begin{figure}[h]
    
    \centering
    \includegraphics[width=0.85\textwidth]{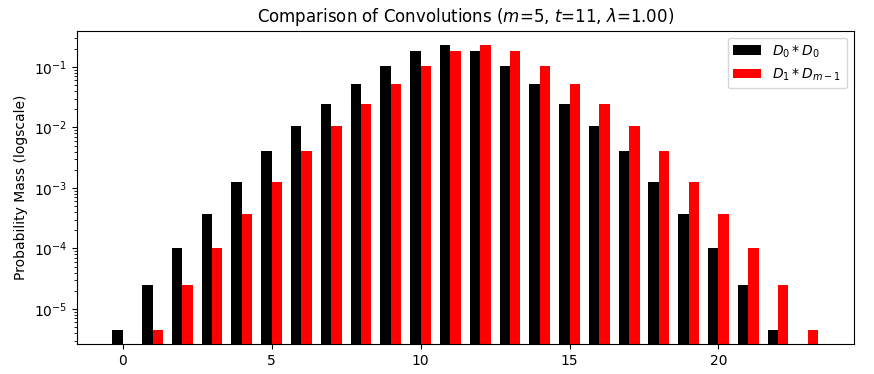}
    \captionsetup{width=0.85\textwidth,font=small}
    \caption{Visualization of two probability mass functions. The black bars characterize the distribution $(S\circ \overline{R}^2)(0,0)$ while the red bars describe $(S\circ \overline{R}^2)(1,m-1)$.}
    
    \label{fig:convolution}
\end{figure}

\paragraph{Correcting the attempt.} But not all hope is lost. In Figure \ref{fig:convolution}, we plot the probability mass functions of the convolutions $\bD_0 * \bD_0$ and $\bD_1 * \bD_{m-1}$. These characterize the number of ones in a sample from $(S\circ \overline{R}^2)(\vec{x})$ and $(S\circ \overline{R}^2)(\vec{x}\,')$, respectively. Observe that the mass functions are similar near the modes and then gracefully diverge from each other as we approach the tails.

The figure suggests the following solution: inflate the tails slightly so that the supports of the two distributions are identical and every element in that support receives approximately the same mass. More precisely, instead of sampling from $\bD_0 * \bD_0$, we sample from a \emph{mixture} between $\bD_0 * \bD_0$ and another distribution $\bB$. To preserve correctness, this mixture should be heavily weighted toward $\bD_0 * \bD_0$. To compensate for that bias, the distribution $\bB$ should place a lot of mass on the elements in the upper tail of $\bD_1 * \bD_{m-1}$. From a symmetric line of thought, $\bB$ should also place much mass on the lower tail elements of $\bD_0 * \bD_0$.

We will construct $\bB$ in a way that eases the inevitable calculations involving the exponential function. We first define $\bL$ to be a ``mirrored'' version of $\bD$: for $u \leq \lfloor t/2 \rfloor $, $\bL[u] := \bD[\lfloor t/2\rfloor -u]$ and otherwise $\bL[u] := \bD[u-\lceil t/2 \rceil]$. This distribution has the same support as $\bD$ but places exponentially more mass on points far from $t/2$ than points near. Now, for all $x\in\{0,1,\dots,m-1\}$ we can construct $\bL_x$ in an analogous way with $\bD_x$: for all $u \in [0,t]$, $\bL_x[x+um] := \bL[u]$,  Finally, we define $\bB[x+um] := \frac{1}{m}\sum_{y=0}^{m-1} \bL_y[x+um] = \frac{1}{m}\cdot \bL[u]$.

Now, we have
\begin{center}
$R(x) :=$ binary vector with length $mt+(m-1)$ \& sum $\eta$, where $\eta \sim (1-p)\cdot \bD_x + p\cdot \bB$
\end{center}

\subsection{Rigorous Analysis}
In this subsection, $S^{d,2,n}_{\eps,q}$ refers to an arbitrary $(\eps,q)$-relaxation of $S^{d,2,n}$.

\begin{thm}
\label{thm:llr-simulation}
Fix any values $\hat{\eps},\hat{q} \in(0,1)$, modulus $m\geq 2$, security parameter $\eps \geq 0$, failure probability $q\in(0,1/2)$ and let $d= mt+(m-1)$. There exist choices for $\lambda,p,t$ such that $S^{d,2,n}_{\eps,q} \circ R^n$ is an $(\eps+\hat{\eps}, q + \hat{q})$-relaxation of $\Sigma^{1,m,n}$.
\end{thm}

\begin{proof}
If we set $p=\hat{q}/n$, correctness follows from a union bound: except with probability $\leq np = \hat{q}$, each user $i$ will sample from the distribution $\bD_{x_i}$. And except with probability $\leq q$, the output of $S^{1,m,n}_{\eps,q}$ consists of $n\cdot k$ bits whose sum is $\sum x_i +vm$ for some positive integer $v$. Thus, the analyzer reports the correct sum modulo $m$ except with probability $\leq q+\hat{q}$.

The remainder of the proof focuses on security. We will describe a simulator $\SIM$ such that, on any input $\vec{x}\in \{0,\dots,m-1\}^n$,
\begin{equation}
\label{eq:llr-sim-objective}
\mathit{LLR}((S^{d,2,n}_{\eps,q}\circ R^n)(\vec{x})),~ \SIM( \Sigma^{1,m,n}(\vec{x}) ) ) \leq \eps + \hat{\eps}    
\end{equation}

Recall that $\eps$-LLR security of intermediary $S^{d,2,n}_{\eps,q}$ implies some simulator $\SIM_S$ where
\begin{equation}
\label{eq:llr-sim-1}
\llr(S^{d,2,n}_{\eps,q}(\vec{y}), \SIM_S( S^{d,2,n}(\vec{y}) ) \leq \eps
\end{equation}
for all inputs $\vec{y}\in (\{0,\dots,m-1\}^k)^{\leq n}$.

Suppose that we constructed $\SIM$ such that
\begin{equation}
\label{eq:llr-sim-2}
    \llr( \SIM_S(S^{d,2,n}(R^n(\vec{x}))), \SIM(\Sigma^{1,m,n}(\vec{x})) ) \leq \hat{\eps}
\end{equation}
Then \eqref{eq:llr-sim-1} and \eqref{eq:llr-sim-2} would together imply \eqref{eq:llr-sim-objective} via the triangle inequality.

Let $\SIM$ be the algorithm that, when given the input $a$, constructs
$$
\vec{y} \gets R^{n}\left( a , 0,\dots, 0 \right)
$$
and then reports $\SIM_S(S(\vec{y}))$. We now prove \eqref{eq:llr-sim-2}:
\begin{align*}
& \llr( \SIM_S(S^{d,2,n}(R^n(\vec{x}))), \SIM(\Sigma^{1,m,n}(\vec{x})) )\\
={}& \llr \paren{ \SIM_S( S^{d,2,n}(R^n(\vec{x}))), \SIM_S \paren{ S^{d,2,n}\paren{ R^n\paren{ \sum_{i\in[n]} x_i \textrm{~mod~} m, 0,\dots,0 } } } } \\
\leq{}& \llr\paren{ S^{d,2,n}(R^n(\vec{x})), S^{d,2,n}\paren{ R^n\paren{ \sum_{i\in[n]} x_i \textrm{~mod~} m, 0,\dots,0 } } } \stepcounter{equation} \label{eq:llr-sim-3} \tag{\theequation}
\end{align*}
The last line follows from the data processing inequality.

We construct the following hybrids, where sums are evaluated modulo $m$:
\begin{align*}
\vec{x} = \vec{x}^{(1)} &:= x_1, x_2, x_3, \dots, x_{n}\\
\vec{x}^{(2)} &:= x_1+x_2, 0, x_3, \dots, x_{n} \\
\vec{x}^{(3)} &:= x_1+x_2+x_3, 0, 0, \dots, x_{n} \\
\dots &\\
\vec{x}^{(n)} &:= \sum x_i, 0, 0, \dots, 0
\end{align*}
A similar set of hybrids previously appeared in work by Ishai, Kushilevitz, Ostrovsky, and Sahai \cite{IKOS06}. Now, the triangle inequality implies
$$
\eqref{eq:llr-sim-3} \leq \sum_{j=1}^{n-1} \llr\left( S_{2,k}(R^n(\vec{x}^{(j)})), S_{2,k}((R^n(\vec{x}^{(j+1)})) \right).
$$
Thus, to complete the proof of \eqref{eq:llr-sim-2}, it will suffice to find $\lambda, t$ where the following holds for every integer $j \in [1, n-1 ]$:
$$
\llr(S^{d,2,n}(R^{n}(\vec{x}^{(j)})), S^{d,2,n}(R^{n}(\vec{x}^{(j+1)}))) ) \leq \hat{\eps} / n.    
$$

We make two key observations. First, $S$ can be equated with a two-step procedure: compute the sum of the bits inside all its input vectors, then produce a sequence of bits of the form $(0,\dots,0,1,\dots,1)$ where the number of ones is the previously computed sum. Second, observe that $x^{(j)}_i = x^{(j+1)}_i$ for all $i\notin \{1,j+1\}$ which implies $R(x^{(j)}_i)$ is distributed identically with $R(x^{(j+1)}_i)$.

Let $G(x_1,x_2)$ be the algorithm that reports the sum of all the bits made by $R^2(x_1,x_2)$. The two preceding observations jointly imply that it will suffice to show
\begin{equation}
\label{eq:llr-sim-4}
\llr(G(x_1,x_2), G(x'_1,x'_2)) \leq \hat{\eps} / n
\end{equation}
for any $x_1+x_2 = x'_1+x'_2 \textrm{~mod~} m$.

For any $z \in \{0,\dots,2k\}$, we decompose the masses placed on $z$ as follows:
\begin{align*}
&\pr{}{G(x_1,x_2) = z}\\
={}& (1-p)^2\cdot \underbrace{(\bD_{x_1} * \bD_{x_2})[z]}_{\alpha_z} + p(1-p) \cdot \underbrace{(\bD_{x_1} * \bB)[z]}_{\beta_z} + p(1-p) \cdot \underbrace{(\bD_{x_2} * \bB)[z]}_{\gamma_z} + p^2 \cdot \underbrace{(\bB*\bB)[z]}_{\delta_z}\\
&\pr{}{G(x'_1,x'_2) = z}\\
={}& (1-p)^2\cdot \underbrace{(\bD_{x'_1} *  \bD_{x'_2})[z]}_{\alpha'_z} + p(1-p) \cdot \underbrace{(\bD_{x'_1} * \bB)[z]}_{\beta'_z} + p(1-p) \cdot \underbrace{(\bD_{x'_2} * \bB)[z]}_{\gamma'_z} + p^2 \cdot \underbrace{(\bB*\bB)[z]}_{\delta_z}
\end{align*}
For each of the terms $\alpha_z, \beta_z, \gamma_z$, we will show it is bounded by a function of the terms $\alpha'_z, \beta'_z, \gamma'_z, \delta_z$ (and vice versa). When combined, these bounds will imply $\llr(G(x_1,x_2), G(x'_1,x'_2)) \leq \hat{\eps} / n$.

\smallskip

\noindent \underline{The $\alpha_z,\alpha'_z$ terms}: We begin with the following fact:
\begin{fact}
\label{fact:mod-math}
For any $x_1,x_2,x'_1,x'_2 \in \{0,1,\dots,m-1\}$, if $x'_1 + x'_2 \textrm{~mod~} m = x_1 + x_2 \textrm{~mod~} m$, then there is some $c \in \{-m,0,+m\}$ such that $x'_1 + x'_2 = x_1 + x_2 + c$.
\end{fact}

One can prove the following from a brief calculation:
\begin{clm}
\label{clm:easy-case}
If $x_1+x_2 = x'_1+x'_2$ then $\alpha'_z=\alpha_z$
\end{clm}

So we will focus on the case where $x'_1 + x'_2 = x_1 + x_2 + m$. The reverse case will hold true by symmetry. The following claim essentially formalizes the pattern observed in Figure \ref{fig:convolution}.

\begin{clm}
\label{clm:bounding-alpha}
If $x'_1 + x'_2 = x_1 + x_2 + m$ and $z=x_1+x_2 + vm$ for integer $v\in[0,t]$, then
\begin{align*}
\alpha_z &\leq e^\lambda \cdot \alpha'_z + \bD[v] \cdot \bD[0]\\
\alpha'_z &\leq e^\lambda \cdot \alpha_z
\end{align*}
If $v\in [t+1,2t+1]$, then
\begin{align*}
\alpha'_z &\leq e^\lambda \cdot \alpha_z + \bD[2t+1-v] \cdot \bD[0]\\
\alpha_z &\leq e^\lambda \cdot \alpha'_z
\end{align*}
Otherwise, $\alpha_z = \alpha'_z$.
\end{clm}

To continue with the proof, we rely on two key insights. First: the term $\alpha'_z$ can be equated with a geometric series. Second: for sufficiently large $v$, the extra additive term $\bD[v]\cdot \bD[0]$ constitutes only a small fraction of that geometric series. These insights allows us to transform the multiplicative-and-additive bound in Claim \ref{clm:bounding-alpha} into a purely multiplicative one. The same holds for $\bD[2t+1-v]\cdot \bD[0]$ and sufficiently small $v$.

\begin{clm}
\label{clm:extra-alpha-term-1}
Fix any $x'_1 + x'_2 = x_1 + x_2 + m$. If $z=x_1+x_2 + vm$ for integer $v \in [1/\lambda,t]$, then $\bD[v] \cdot \bD[0] \leq 3\lambda \cdot \alpha'_z$.  If $v \in [t+1,2t+1-1/\lambda]$, then $\bD[2t+1-v]\cdot \bD[0] \leq 3\lambda \cdot \alpha_z$.
\end{clm}

For small values of $v$, we claim $\bD[v]$---a value in the tail of the discrete Laplace distribution---makes up only a small fraction of the geometric sequence equivalent to $\delta_z = (\bB*\bB)[z]$. This comes from the fact that on the extreme values of $v$, we engineered $\bB$ to place exponentially more mass than the discrete Laplace distribution. Via a symmetric argument, the same is true for large values of $v$ and $\bD[2t+1-v]$.

\begin{clm}
\label{clm:extra-alpha-term-2}
Fix any $x'_1 + x'_2 = x_1 + x_2 + m$ and any odd $t \geq 3 + \frac{4}{\lambda} + \frac{2}{\lambda}\cdot \ln \frac{m^2}{\lambda\cdot p^2}$. If $z=x_1+x_2 + vm$ for integer $v \notin [1/\lambda, 2t+1-1/\lambda]$, then $\bD[\min(v,2t+1-v)] \cdot \bD[0] \leq \lambda \cdot p^2 \cdot \delta_z$
\end{clm}

We will prove the four preceding claims in Appendix \ref{sec:alpha}. Combined, they imply the following corollary:
\begin{coro}
\label{coro:alpha}
Fix any odd $t \geq 3 + \frac{4}{\lambda} + \frac{2}{\lambda}\cdot \ln \frac{m^2}{\lambda\cdot p^2}$. For any $x_1,x_2,x'_1,x'_2 \in \{0,1,\dots,m-1\}$ where $x'_1 + x'_2 \textrm{~mod~} m = x_1 + x_2 \textrm{~mod~} m$,
\begin{align*}
    \alpha_z &\leq e^{4\lambda}\cdot \alpha'_z + \lambda\cdot p^2 \cdot \delta_z\\
    \alpha'_z &\leq e^{4\lambda}\cdot \alpha_z + \lambda\cdot p^2 \cdot \delta_z
\end{align*}
\end{coro}

\noindent \underline{The $\beta_z,\beta'_z$ and $\gamma_z,\gamma'_z$ terms}. Notice that these pairs of terms are symmetric, so it will suffice to prove the following generic claim:

\begin{clm}
\label{clm:bounding-beta}
Fix any odd $t \geq 3 + \frac{4}{\lambda} + \frac{2}{\lambda}\cdot \ln \frac{m}{\lambda\cdot p}$. For any $x,x'\in\{0,1,\dots,m-1\}$, $$(\bD_x * \bB)[z] \leq e^{4\lambda}\cdot (\bD_{x'} * \bB)[z] + \lambda\cdot p\cdot (\bB * \bB)[z].$$
\end{clm}

The techniques to prove this claim are essentially identical to the proof of Corollary \ref{coro:alpha}. Refer to Appendix \ref{sec:beta} for the proof.

\smallskip

\noindent \underline{Wrapping up:} Taking Corollary \ref{coro:alpha} and Claim \ref{clm:bounding-beta} together, we have
\begin{align*}
    &(1-p)^2\cdot \alpha_z+ p(1-p)\cdot \beta_z+ p(1-p)\cdot \gamma_z+ p^2\cdot \delta_z\\
    \leq{}& e^{4\lambda} \cdot (1-p)^2\cdot \alpha'_z + p(1-p)\cdot \beta_z+ p(1-p)\cdot \gamma_z+ (1+\lambda)\cdot p^2\cdot \delta_z \\
    \leq{}& e^{4\lambda} \cdot (1-p)^2\cdot \alpha'_z + p(1-p)\cdot e^{4\lambda}\beta'_z+ p(1-p)\cdot e^{4\lambda} \gamma'_z+ (1+3\lambda)\cdot p^2\cdot \delta_z \\
    \leq{}& e^{4\lambda}\cdot \left[ (1-p)^2\cdot \alpha'_z+ p(1-p)\cdot \beta'_z+ p(1-p)\cdot \gamma'_z+ p^2\cdot \delta_z \right]
\end{align*}
The reverse inequality also holds:
\begin{align*}
    &(1-p)^2\cdot \alpha'_z+ p(1-p)\cdot \beta'_z+ p(1-p)\cdot \gamma'_z+ p^2\cdot\delta_z\\
    \leq{}& e^{4\lambda}\cdot \left[ (1-p)^2\cdot \alpha'_z+ p(1-p)\cdot \beta'_z+ p(1-p)\cdot \gamma'_z+ p^2\cdot \delta_z \right]
\end{align*}
We simply substitute $\lambda=\hat{\eps}/4n$ to conclude that, for any $z$,
$$
\ln \max \left( \frac{\pr{}{G(x_1,x_2) = z}}{\pr{}{G(x'_1,x'_2) = z}}, \frac{\pr{}{G(x'_1,x'_2) = z}}{\pr{}{G(x_1,x_2) = z}} \right) \leq \hat{\eps}/n
$$
which is equivalent to the desired inequality \eqref{eq:llr-sim-4}.
\end{proof}

\paragraph{Extending to larger dimensions.} The above construction lets us instantiate $\Sigma^{d=1,m,n}$. Here, we sketch how to extend the construction to $d>1$. The main idea is to execute the simulation $d$ times in parallel, using labels to disambiguate messages.

When given the data vector of the $i$-th user $(x_{i,1},\dots,x_{i,d})$, the new randomizer runs $R$ on each $x_{i,j}$. For every bit produced by $R(x_{i,j})$ the new randomizer constructs a tuple that pairs $j$ with the bit. The messages of the new protocol are these labeled bits. The labels allow the new analyzer to run the old function $A$ once for every $j\in[d]$ on the corresponding bits.

\paragraph{Attacks against the construction.} In our proof, we showed that our relaxed aggregator is secure in the sense that the output does not contain much more information than the sum of the inputs. But we intend to use this relaxed aggregator inside a \emph{robustly differentially private} protocol. In that setting, an adversary not only views the intermediary's output but also controls a fraction of the users (Definition \ref{defn:robust-dp}). Because our construction of the relaxed aggregator relies on the participation of users, the security guarantee should likewise be robust to corrupt users.

Fortunately, our arguments easily extend to that case. Without loss of generality, suppose the adversary controls the users indexed by $C = [c]$ and user $i\in C$ sends some adversarially chosen $y_i$. We claim the adversary cannot learn much more than the sum of \emph{honest} user inputs, regardless of the choice of $y_1,\dots,y_c$. Formally, our objective is to construct a new simulator $\SIM'$ such that the following variant of \eqref{eq:llr-sim-2} holds: $$\llr( \SIM_S(S^{d,2,n}(y_1,\dots,y_c,R^{n-c}(\vec{x}_{\overline{C}}))), \SIM'(y_1,\dots,y_c,\Sigma^{1,m,n}(\vec{x}_{\overline{C}})) ) \leq \hat{\eps} $$
The new simulator, on input $y_1,\dots,y_c,a$, constructs $\vec{y} \gets (y_1,\dots,y_c,R(a),R(0),\dots,R(0))$, and then reports $\SIM_S(S^{d,2,n}(\vec{y}))$. The rest of the proof proceeds in much the same manner as before, except that we use $n-c$ hybrids instead of $n$ hybrids.

\subsection{An Impossibility Result for the Shuffler}

The preceding construction implies that the shuffler $S^{d,m,n}$ can be used for an $(\eps,q)$-relaxed aggregator, where $\eps >0$. It is natural to ask if we can go even further: can the shuffler be used for an $(0,q)$-implementation? Is perfect security possible? In Appendix \ref{apdx:sim-perfect-mod}, we show it is not.

The proof has the following structure. First, we argue that a randomizer which sends binary vectors is powerful enough to simulate any other randomizer. Then, we make the observation that a shuffled set of binary values is equivalent to the sum of their values. This lets us reason about the moment generating function of the distribution of ones.

\begin{clm}
\label{clm:binary-reduction}
For any sequence of randomizers $\vec{R}$ where $R_i : \cX \to \{0,\dots,m-1\}^k$, there exists an integer $k'$, a sequence of randomizers $\vec{R}\,'$ where $R'_i: \cX\to \zo^{k'}$ and a function \POST~ such that $(\POST \circ S^{d,2,n'} \circ \vec{R}\,')(\vec{x})$ is identically distributed with $(S^{d,m,n}\circ \vec{R})(\vec{x})$ for all inputs $\vec{x}\in\cX^n$.
\end{clm}

\begin{clm}
\label{clm:perfect-lower-bound}
Fix any number of users $n \geq 2$, message complexity $k \in \N$, and modulus $m \geq 2$. If $(S^{d,2,n} \circ \vec{R})$ is a $(0,q)$-relaxation of $\Sigma^{1,m,n}$, then $q > 1-\frac{1}{m}$.
\end{clm}

\section{Pure DP Sums from Secure Aggregation}
\label{sec:bounded-sum}

In this section, we describe a protocol which adds values in the interval $[0,1]$ while satisfying $\eps$-differential privacy. The expected error due to privacy will be $O(1/\eps)$. As explained in the introduction, our construction swaps out statistically secure aggregation in Balle et al.'s protocol \cite{BBGN19-2} with LLR-secure aggregation. The change in security guarantee allows us to prove pure differential privacy instead of approximate differential privacy.

We give a high-level overview of the protocol. First, each user uses randomized rounding to map their datum $x_i\in[0,1]$ to an integer $\phi_i \in [g]$. Then they add a small amount of noise to that encoded value. This noise is drawn in such a way that the aggregate noise is the sum of two samples from a discrete Laplace distribution. This ensures robust differential privacy, since one sample from the discrete Laplace distribution will be added whenever half the users are honest.

With respect to accuracy, we note that there are two sources of error: randomized rounding and privacy noise. A sufficiently large encoding size $g$ will make the error from randomized rounding a lower-order term. Meanwhile, a large value of modulus $m$ ensures that the noise from privacy is very unlikely to cause overflow. And noise from privacy will likely have magnitude $O(1/\eps)$.

Although our description of the protocol uses $\Sigma^{1,m,n}_{\eps,q}$---an arbitrary relaxation of the aggregator $\Sigma^{1,m,n}$---we stress that this can be instantiated with a relaxed shuffler (Section \ref{sec:sim-mod}).

\begin{myalgorithm}
\caption{A local randomizer $R$ for sums of $[0,1]$ values}
    \label{alg:summation-randomizer}
    
    \KwIn{Data $x_i \in[0,1]$; parameters $m,g\in \N, ~\lambda \in (0,1)$}
    
    Encode $x_i$ as $\phi_i \gets \lfloor x_i \cdot g \rfloor + \Ber(x_i \cdot g - \lfloor x_i \cdot g \rfloor)$
    
    Sample two independent samples $\eta^{+}_i,\eta^{-}_i \sim \Polya(2/n,\lambda)$
    
    Compute $y_i \gets \phi_i + \eta^{+}_i - \eta^{-}_i ~\mathrm{mod}~ m$
    
    \Return{$y_i$}
\end{myalgorithm}






\begin{myalgorithm}
\caption{An analyzer function $A$ for sums of $[0,1]$ values}
\label{alg:summation-analyzer}

\KwIn{Value $y\in\{0,\dots,m-1\}$; parameters $m,g,\tau \in \N$}
\KwOut{$z\in\R$, an estimate of the bounded sum}

\If{$y > ng+2\tau$}{ $z \gets (y-m)/g$ }
\Else{$z \gets y/g$}

\Return{$z$}
\end{myalgorithm}

\begin{thm}
\label{thm:bounded-sum}
Let $P=(R,\Sigma^{1,m,n}_{\hat{\eps},\hat{q}},A)$ be the protocol where $R,A$ are specified by Algorithms \ref{alg:summation-randomizer} and \ref{alg:summation-analyzer}, respectively. Fix any $\eps,q \in (0,1)$. If $\lambda \gets \exp(-\eps/g)$ then $P$ is $(2\hat{\eps} + \eps)$-robustly differentially private. Moreover, there are choices of $g,\tau,m$ such that $P(\vec{x})$ is within $O(\frac{1}{\eps} \log\frac{1}{q})$ of $\sum x_i$ with probability $\geq 1-(\hat{q}+3q)$.
\end{thm}

\begin{proof}
\underline{Robust Privacy}: Our goal is to prove the following
$$
\llr(\view_T(P,\vec{x}), \view_T(P,\vec{x}\,')) \leq 2\hat{\eps} + \eps
$$
for all attacks $T=(C,W)$ where $|C|\leq n/2$ and any input vectors $\vec{x},\vec{x}\,'$ which differ on one index. For neatness, we will assume without loss of generality that $\overline{C}$ is a prefix of $1,2,3,\dots,n$. Now, the left-hand side of the above expands to
\begin{align*}
& \llr(\view_T(P,\vec{x}), \view_T(P,\vec{x}\,')) \\
    ={}& \llr\bigg( \Sigma^{1,m,n}_{\hat{\eps},\hat{q}} \left(R(x_1),\dots,R(x_{|\overline{C}|}), w_{|\overline{C}|+1} , \dots, w_n \right),\\
        & \Sigma^{1,m,n}_{\hat{\eps},\hat{q}} \left(R(x'_1),\dots,R(x'_{|\overline{C}|}), w_{|\overline{C}|+1} , \dots, w_n \right) \bigg)\\
    \leq{}& 2\hat{\eps} + \llr\Bigg( \SIM\paren{ \sum_{i\notin C} R(x_i) + \sum_{i \in C} w_i \textrm{~mod~} m },\\
        & \SIM\paren{ \sum_{i\notin C} R(x'_i) + \sum_{i \in C} w_i \textrm{~mod~} m } \Bigg) \tag{Security of $\Sigma^{1,m,n}_{\hat{\eps},\hat{q}}$}\\
    \leq{}& 2\hat{\eps} +\llr\left(\sum_{i\notin C} R(x_i) \textrm{~mod~} m, \sum_{i\notin C} R(x'_i) \textrm{~mod~} m \right)
\end{align*}
The last step follows from the data processing inequality.

In order to conclude the proof, we need to bound the distance in the above inequality by $\eps$. To do so, observe that $H$ (pseudocode in Algorithm \ref{alg:central-summation}) simulates the computation in question. This implies
$$
\llr\paren{\sum_{i\notin C} R(x_i) \textrm{~mod~} m, \sum_{i\notin C} R(x'_i) \textrm{~mod~} m } \leq \llr\paren{H(\vec{x}_{\overline{C}}) \textrm{~mod~} m, H(\vec{x}\,'_{\overline{C}}) \textrm{~mod~} m }
$$
Thus, it would suffice to prove $H$ is $\eps$-differentialy private on $\geq n/2$ inputs.

\begin{myalgorithm}
\caption{An algorithm $H$ for sums of $[0,1]$ values}
\label{alg:central-summation}

\KwIn{$\vec{x}\in [0,1]^*$, a vector of sensitive inputs; parameters $g\in \N, ~\lambda \in (0,1)$}
\KwOut{$v \in \Z$, an estimate of the sum of the inputs}

\For{each index $i$ in input}{

    Encode $x_i$ as $\phi_i \gets \lfloor x_i \cdot g \rfloor + \Ber(x_i \cdot g - \lfloor x_i \cdot g \rfloor)$

    Sample two independent samples $\eta^{+}_i,\eta^{-}_i \sim \Polya(2/n,\lambda)$
    
}

\Return{$v\gets \sum_i (\phi_i + \eta^{+}_i-\eta^{-}_i)$}
\end{myalgorithm}

Recall that the discrete Laplace distribution is infinitely divisible: if there were exactly $n/2$ inputs to $H$, the aggregate noise added to $\sum \phi_i$ would be drawn from $\DLap(\lambda)$ (Fact \ref{fact:polya-dlap}). Moreover, setting $\lambda \gets e^{-\eps/g}$ would ensure $\eps$-differential privacy because the sensitivity of $\sum \phi_i$ is $g$ (Theorem \ref{thm:lap-mech}). When there are more than $n/2$ inputs, we can interpret the ``surplus noise'' as post-processing and privacy would carry over.

\medskip

\underline{Accuracy}: We will assign $g\gets \lceil \eps\cdot \sqrt{n} \rceil$, $\tau \gets \lceil \frac{g}{\eps}\ln \frac{2}{q} \rceil$, and $m\gets ng+ 4\tau$.

We first show that the estimate reported by the analyzer is a good approximation of the quantity $\frac{1}{g}\sum_{i=1}^n \phi_i$:
\begin{equation}
\label{eq:bounded-sum-1}
\pr{}{\left| A(y)-\frac{1}{g}\sum_{i=1}^n \phi_i \right| > \frac{2\tau}{g}} \leq \hat{q}+2q
\end{equation}
Then we argue that $\frac{1}{g}\sum_{i=1}^n \phi_i$ is a good approximation of the underlying sum:
\begin{equation}
\label{eq:bounded-sum-2}
\pr{}{\left| \frac{1}{g}\cdot \sum_{i\in [n]} \phi_i - \sum_{i\in [n]} x_i \right|>\frac{1}{\eps} \sqrt{\ln \frac{2}{q}}} \leq q.    
\end{equation}
The claim follows from the triangle inequality and a union bound.

\medskip

To prove \eqref{eq:bounded-sum-1}, we rely on two events that hold with high probability. First, $\Sigma^{1,m,n}_{\hat{\eps},\hat{q}}$ correctly performs modular sums except with probability $\leq \hat{q}$. This is a modeling assumption. Next, the magnitude of privacy noise is unlikely to exceed $2\tau$:
\begin{equation}
\label{eq:bounded-sum-3}
\pr{}{\left| \sum_{i\in[n]} \eta^{+}_i - \eta^{-}_i \right| > 2\tau} \leq 2q.
\end{equation}
To prove the above inequality, we use the fact that $\sum_{i=1}^{n/2} \eta^{+}_i - \eta^{-}_i$ and $\sum_{i=n/2 + 1}^{n} \eta^{+}_i - \eta^{-}_i$ are each distributed as $\DLap(\lambda)$. So the probability that $\sum_{i=1}^{n/2} \eta^{+}_i - \eta^{-}_i =v$ is $\propto \lambda^{|v|} = e^{-|v|\cdot \eps/g}$. A fairly straightforward calculation shows that $|\sum_{i=1}^{n/2} \eta^{+}_i - \eta^{-}_i|$ exceeds $\tau$ with probability at most $q$. The same can be said for $\sum_{i=n/2 + 1}^{n} \eta^{+}_i - \eta^{-}_i$. So \eqref{eq:bounded-sum-3} follows from a union bound.

We bring the focus back to \eqref{eq:bounded-sum-1}. From \eqref{eq:bounded-sum-3}, we know that the noised sum $\sum_{i=1}^n \phi_i+ \eta^{+}_i - \eta^{-}_i$ lies in the interval $[\sum_{i=1}^n \phi_i - 2\tau, \sum_{i=1}^n \phi_i + 2\tau]$ except with probability $2q$. Meanwhile, by the correctness of the aggregator, the analyzer's input is
\begin{align*}
y &= \left( \sum_{i=1}^n y_i \right) \textrm{~mod~} m \\
    &= \left( \sum_{i=1}^n \phi_i+\eta^{+}_i - \eta^{-}_i \right) \textrm{~mod~} m    
\end{align*}
except with probability $q$. Conditioned on these high-probability events, we will analyze $y$ via case analysis.

\emph{Case 1}: $y > ng+2\tau$. Because $m$ is sufficiently large and $\forall i ~ \phi_i \leq g$, observe that $y > ng+2\tau$ must mean that $\sum_{i=1}^n \phi_i+\eta_i \in [-2\tau,0]$. Therefore,
\begin{align*}
y-m ={}& \left( \left( \sum_{i=1}^n \phi_i+\eta^{+}_i + \eta^{-}_i \right) \textrm{~mod~} m \right)- m\\
={}& \left(m+ \sum_{i=1}^n \phi_i+\eta^{+}_i + \eta^{-}_i \right)- m\\
={}& \sum_{i=1}^n \phi_i+\eta^{+}_i + \eta^{-}_i\\
\in{}& \left[ \sum_{i=1}^n \phi_i - 2\tau, \sum_{i=1}^n \phi_i + 2\tau \right]\\
\therefore \frac{y-m}{g} \in{}& \left[ \frac{1}{g}\sum_{i=1}^n \phi_i - \frac{2\tau}{g}, \frac{1}{g}\sum_{i=1}^n \phi_i + \frac{2\tau}{g} \right]
\end{align*}

\emph{Case 2}: $y\leq ng+2\tau$. Here, the noised sum $\sum_{i=1}^n \phi_i+\eta^{+}_i + \eta^{-}_i$ lies in the interval $[0, ng+2\tau]$ so $y/g \in [ \frac{1}{g}\sum_{i=1}^n \phi_i - \frac{2\tau}{g}, \frac{1}{g}\sum_{i=1}^n \phi_i + \frac{2\tau}{g}]$.

In both cases, we have shown that the estimate reported by the analyzer is within $2\tau/g= O(\frac{1}{\eps}\log \frac{1}{q})$ of $\frac{1}{g}\sum_{i=1}^n \phi_i$.

It now remains to prove \eqref{eq:bounded-sum-2}, which is that $\frac{1}{g}\sum_{i=1}^n \phi_i$ is close to the sum of the inputs $\sum_{i\in [n]} x_i$. By construction, the expected value of $\phi_i / g$ is $x_i$ and it lies in the range $[0,1]$. Its variance is 
\begin{align*}
\var{}{\frac{1}{g}\cdot \phi_i} &= \frac{1}{g^2}\cdot \var{}{\Ber(x_i g - \lfloor x_i g\rfloor)} \\
&\leq \frac{1}{g^2}\cdot \var{}{\Ber(1/2)}\\
&= \frac{1}{4g^2} = \frac{1}{4\eps^2 n}
\end{align*}
We can now invoke a Chernoff bound: except with probability $q$, the distance between $\frac{1}{g}\sum_{i\in[n]} \phi_i$ and $\sum_{i\in [n]} x_i$ is at most $2\sqrt{n\cdot \var{}{\frac{1}{g}\cdot \phi_i} \cdot \ln \frac{2}{q} } = \frac{1}{\eps} \sqrt{\ln \frac{2}{q}}$.
\end{proof}

\medskip

In the special case where data is not real-valued but binary, note that we can eliminate error due to randomized rounding. This allows us to give a fairly clean characterization of the noise distribution:
\begin{clm}[Noise added to Binary Data]
\label{clm:binary-sum-noise}
Suppose we assign $g\gets 1$ but assign $m,\tau,\lambda$ in the same fashion as Theorem \ref{thm:bounded-sum}. If $\vec{x} \in \zo^n$ and we compute $z\gets P(\vec{x})$, then there exists a distribution $D$ and value $\gamma \in (0,\hat{q}+2q]$ such that the error $z - \sum_{i\in [n]} x_i$ is distributed as the mixture
$$
\gamma \cdot D + (1-\gamma)\cdot (\DLap_{\tau}(\lambda) * \DLap_{\tau}(\lambda)).
$$
\end{clm}
\begin{proof}
In this setting, the quantity $\sum \phi_i$ is \emph{exactly} $\sum x_i$ which means \eqref{eq:bounded-sum-1} is the only error bound of relevance. And recall that the steps taken to argue \eqref{eq:bounded-sum-1} includes an argument that the random variables $\sum_{i=1}^{n/2} \eta^{+}_i-\eta^{-}_i$ and $\sum_{i=n/2+1}^{n} \eta^{+}_i-\eta^{-}_i$ are each drawn from $\DLap(\lambda)$. As a consequence, the overall noise is drawn from a convolution between two copies of $\DLap(\lambda)$.

$\DLap(\lambda)$ can be expressed as a mixture between $\DLap_\tau(\lambda)$ and some other distribution over the integers with magnitude larger than $\tau$. The weight placed on the second distribution is $\leq q$, by construction of $\tau$. Using this decomposition, we can express $\DLap(\lambda) * \DLap(\lambda)$ as a mixture between $\DLap_\tau(\lambda) * \DLap_\tau(\lambda)$ and some other distribution over integers. The weight placed on the second distribution is $\leq 2q$, by a union bound.

In the event that the noise is sampled from $\DLap_\tau(\lambda) * \DLap_\tau(\lambda)$, our previous case analysis implies that the analyzer function exactly recovers the noised sum. That is, $A(y)$ reports $\sum_{i=1}^n \phi_i+\eta^{+}_i + \eta^{-}_i$ whenever the intermediary $\Sigma^{1,m,n}_{\hat{\eps},\hat{q}}$ correctly computes modular sums (which occurs with probability $\geq 1-\hat{q}$).
\end{proof}

\section{Pure DP Uniformity Testing from Sums}

In this section, we show how to perform uniformity testing while satisfying pure differential privacy. Our core protocol follows a template established by Amin, Joseph, and Mao \cite{AJM20} in the online model, which later found use in the shuffle model by Balcer, Cheu, Joseph, and Mao \cite{BCJM21}. The latter work only achieved approximate differential privacy, while we achieve pure differential privacy in both the aggregation and shuffle models. This is achieved with our counting protocol in the aggregation model (Section \ref{sec:bounded-sum}) and our simulation of an aggregator with a shuffler (Section \ref{sec:sim-mod}).

\medskip

We begin by defining the uniformity testing problem. We will use $\bD_{[d]}$ to denote the uniform distribution over the universe $[d]$; we drop the subscript if the universe size is clear. We model the number of users $n$ as a Poisson random variable $n \sim \Pois(N)$. We assume that each user $i$ samples their datum $x_i$ independently from an unknown distribution $\bD$ over the integers $[d]$. The goal is to use the samples to determine if $\bD$ is the uniform distribution (which we denote $\bD=\bU$) or if it is $\alpha$-far from uniform in statistical distance, for some parameter $\alpha$. Formally,

\begin{defn}
\label{defn:ut}
A protocol $P$ solves $\alpha$-uniformity testing with sample complexity $N^*$ if the following holds for any $\bD$ over $[d]$ and any $N\geq N^*$. On input $\vec{x}\sim \bD^n$ (where $n\sim \Pois(N)$),
\begin{itemize}
    \item $P$ reports ``not uniform'' with probability at most $1/3$ when $\bD=\bU$.
    \item $P$ reports ``not uniform'' with probability at least $2/3$ when $\SD(\bD,\bU) >\alpha$.
\end{itemize}
The randomness of sampling as well as of the protocol itself are factored into the probabilities.
\end{defn}

\begin{rem}
\label{rem:ut}
To prove that a protocol $P$ satisfies the above condition, it will suffice to prove $P$ satisfies a mildly relaxed variant. Specifically, suppose there is a protocol that follows Definition \ref{defn:ut} except that the probabilities are $t,t'$ for $t'-t=\Omega(1)$ instead of $1/3,2/3$. Then there is another protocol whose sample complexity is $O(N^*)$ instead of $N^*$: execute the original protocol $O(1/(t'-t)^2) = O(1)$ times, with $\Pois(N^*)$ new samples each time. When $\bD=\bU$ (resp. when $\norm{\bD-\bU}_{\TV} >\alpha$), a Chernoff bound implies that the fraction of times the protocol reports ``not uniform'' is $>(t+t')/2$ with probability at most 1/3 (resp. at least 2/3).
\end{rem}

The prior works by Balcer et al. \cite{BCJM21} and Cheu \cite{Cheu21} prove a lower bound on the sample complexity of any symmetric robustly private shuffle protocol. But the arguments can be easily extended to the asymmetric case, as well as robustly private aggregation protocols.
\begin{thm}
\label{thm:ut-lower}
Let $P$ be a protocol using either a relaxed shuffler or a relaxed aggregator with error $q=O(1)$. If $P$ is $\eps$-robustly private and solves $\alpha$-uniformity testing, then its sample complexity is $$N^*= \Omega \left( \frac{d^{2/3}}{\alpha^{4/3} \eps^{2/3}} + \frac{d^{1/2}}{\alpha^2} + \frac{1}{\alpha\eps} \right).$$
\end{thm}
Our goal is to match this lower bound.

\subsection{A Preliminary Protocol}

As stated previously, we follow a template established in prior work. We compute a private histogram of user data and then compute a noised chi-squared test statistic. The private histogram is obtained via our counting protocol (Section \ref{sec:bounded-sum}).

\begin{myalgorithm}
\caption{A local randomizer for uniformity testing}
\label{alg:ut-randomizer}

\KwIn{Data $x_i \in [d]$; parameters $m\in \N$, $\lambda \in (0,1)$}
\KwOut{Message $y_i \in \{0,\dots,m\}^d$}

\For{$j\in[d]$}{
    Let $\hat{x}_{i,j} \gets \ind{x_i=j}$
    
    Compute $y_{i,j}$ by running Algorithm \ref{alg:summation-randomizer} on $\hat{x}_{i,j}$ with parameters $g\gets 1$, $m$, and $\lambda$
}

\Return{$y_i\gets(y_{i,1},\dots,y_{i,d})$}
\end{myalgorithm}

\begin{myalgorithm}
\caption{An analyzer function for uniformity testing}
\label{alg:ut-analyzer}

\KwIn{Vector $y \in \{0,\dots,m\}^d$; parameters $\ell \in \R$, $m,\tau \in \N$, $\lambda \in (0,1)$}
\KwOut{Either ``uniform'' or ``not uniform''}

\For{$j\in[d]$}{
    \If{$n>0$}{
        Compute $\hat{c}_j$ by running Algorithm \ref{alg:summation-analyzer} on $y_{j}$ with param. $g\gets 1$, $m$, $\tau$, \& $\lambda$
    }
    \Else{
        \tcc{Ensure the added noise is same for all $n$}
        $\hat{c}_j \sim \DLap_\tau(\lambda) * \DLap_\tau(\lambda)$
    }
}

Compute test statistic $Z' \gets \frac{d}{N} \sum_{j\in[d]} (\hat{c}_j - \frac{N}{d})^2 - \hat{c}_j$

\Return{``uniform'' if $Z' \leq \ell$, otherwise ``not uniform''}
\end{myalgorithm}

\begin{thm}
\label{thm:ut-prelim}
Let $P=(R,\Sigma^{d,m,n}_{\hat{\eps},1/200},A)$ be the protocol specified by Algorithms \ref{alg:ut-randomizer} and \ref{alg:ut-analyzer}. Fix any $\eps \in (0,1)$. If $\lambda \gets \exp(-\eps)$, then $P$ is $(4\hat{\eps} + 2\eps)$-robustly differentially private. Moreover, there are choices of $\ell,m,\tau$ such that the sample complexity is $$N^* = O\left( \frac{d^{3/4}}{\alpha\eps} + \frac{d^{2/3}}{\alpha^{4/3}\eps^{2/3}} + \frac{d^{1/2}}{\alpha^2}\right)$$
\end{thm}

Because the techniques are borrowed from prior work, we defer a rigorous proof to Appendix \ref{apdx:ut}. The protocol uses the summation protocol from Section \ref{sec:bounded-sum} to compute a noised histogram of user values. Privacy is immediate from two-fold non-adaptive composition. To derive the sample complexity bound, we first use Claim \ref{clm:binary-sum-noise} to express the noise in the statistic $Z'$ in terms of independent samples from $\DLap_\tau(\lambda)$. Then we use bounds on the moments of $\DLap_\tau(\lambda)$ to show that (1) $Z'$ is likely smaller than the threshold $\ell$ when the protocol is run on data drawn from $\bU$ and (2) $Z'$ is likely larger than $\ell$ when the distribution is $\alpha$-far from uniform.

We remark that the testing template is general enough so that we could have instead used the counting protocol by Ghazi et al. \cite{GGK+20}, which also satisfies pure differential privacy (in the shuffle model). But certain steps we take in our analysis require that the noise added to the counts is symmetrically distributed and independent of user data. These properties are absent from the noise distribution in Ghazi et al.'s protocol, but they are present in truncated discrete Laplace distribution. Moreover, the Ghazi et al. protocol estimates counts with error $O(1/\eps^{3/2})$ instead of $O(1/\eps)$ so that we would have suboptimal dependence on $\eps$ in the sample complexity.

\subsection{The Final Protocol}
The sample complexity bound achieved in Theorem \ref{thm:ut-prelim} has a worse dependence on $d$ than the lower bound in Theorem \ref{thm:ut-lower}. Nevertheless, we show how to use it as a building block of a new protocol whose sample complexity is much closer to the lower bound. Specifically, we manage to reduce the $d^{3/4}/\alpha \eps$ term to a $d^{1/2}/\alpha \eps$ term.

We first introduce some notation. Let \texttt{Coarsen} be the function which takes as input an integer $j\in[d]$ and a partition $G$ of $[d]$ into groups $G_1,\dots,G_{\hat{d}}$ then outputs the integer $\hat{j}\in[\hat{d}]$ such that $j \in G_{\hat{j}}$. Now, let $\bD_G$ be the distribution over $[\hat{d}]$ induced by sampling $j\sim \bD$ and then running \texttt{Coarsen}$(j,G)$. Observe that $\pr{}{\bD_G = \hat{j}}=\sum_{j \in G_{\hat{j}}} \pr{}{\bD=j}$.

We are now ready to give a high level sketch of the final protocol. It first samples a random partition $G$ of $[d]$ and applies the \texttt{Coarsen} function to each user's datum. Then it runs the preliminary tester on the coarsened data (drawn from $\bD_G$). Because the universe is smaller, the sample complexity is reduced. However, the distance from the uniform distribution $\bU$ could also be affected. To place a bound on the change, we leverage the following technical lemma found in work by Acharya, Canonne, Han, Sun, and Tyagi \cite{ACHST20} and  Amin et al. \cite{AJM20}.

\begin{lem}[Domain Compression \cite{ACHST20,AJM20}]
\label{lem:compression}
Let $\bD$ be a distribution over $[d]$. If $G$ is a uniformly random partition of $[d]$ into $\hat{d}<d$ groups of equal size, then with probability $\geq 1/954$ over $G$,
$$
\SD(\bD_G, \bU_{[\hat{d}]} ) \geq \frac{\sqrt{\hat{d}}}{477 \sqrt{10 d}} \cdot \SD( \bD, \bU_{[d]}).
$$
\end{lem}

Refer to Algorithms \ref{alg:ut-final-randomizer} and \ref{alg:ut-final-analyzer} for pseudocode of the final local randomizer and analyzer function.

\begin{myalgorithm}
\caption{A local randomizer for uniformity testing}
\label{alg:ut-final-randomizer}

\KwIn{Data $x_i \in [d]$; parameters $m\in \N$, $\lambda \in (0,1)$}
\KwOut{Message $y_i \in \{0,\dots,m\}^{\hat{d}}$}

Obtain uniformly random $G$ from public randomness

Compute $\hat{x}_i \gets $ \texttt{Coarsen}$(x_i, G)$

\Return{$y_i\gets$ Algorithm \ref{alg:ut-randomizer} run on $\hat{x}_i$ (with parameters $m,\lambda$)}
\end{myalgorithm}

\begin{myalgorithm}
\caption{An analyzer function for uniformity testing}
\label{alg:ut-final-analyzer}

\KwIn{Vector $y \in \{0,\dots,m\}^{\hat{d}}$; parameters $\ell \in \R$, $m,\tau \in \N$, $\lambda \in (0,1)$}
\KwOut{Either ``uniform'' or ``not uniform''}

\Return{output of Algorithm \ref{alg:ut-analyzer} run on $y$ (with universe $\hat{d}$ and parameters $\ell,m,\tau,\lambda$)}
\end{myalgorithm}

\begin{thm}
\label{thm:ut-final}
Let $P=(R,\Sigma^{\hat{d},m,n}_{\hat{\eps},1/200},A)$ be the protocol specified by Algorithms \ref{alg:ut-final-randomizer} and \ref{alg:ut-final-analyzer}. Fix any $\eps \in (0,1)$. If $\lambda \gets \exp(-\eps)$, then $P$ is $(4\hat{\eps} + 2\eps)$-robustly differentially private. Moreover, there are choices of $\hat{d}, \ell, m, \tau$ such that the sample complexity is $$N^* = O\left( \frac{d^{2/3}}{\alpha^{4/3}\eps^{2/3}} + \frac{d^{1/2}}{\alpha\eps} + \frac{d^{1/2}}{\alpha^2} \right)$$
\end{thm}

Just like the previous subsection, we defer the proof to Appendix \ref{apdx:ut}.

\section{Conclusion and Open Questions}
We have shown the feasibility of optimal pure differential privacy in the shuffle and aggregation models. We implement an aggregator with a shuffler in a way that allows simulation of the discrete Laplace mechanism without paying a statistical distance parameter. The summation protocol lets us perform uniformity testing with a sample complexity that has the optimal dependence on dimension $d$.

There are several questions that remain open. For example, is there a more efficient $(\eps+\hat{\eps},q+\hat{q})$-relaxation of the aggregator than the one in Section \ref{sec:sim-mod}? Our randomizer sends a binary vector whose length is $mt= \tilde{O}(m/\lambda) = \tilde{O}(mn/\hat{\eps})$ where $m$ and $n$ are the modulus and number of users respectively. In contrast, the state-of-the-art analysis by Balle, Bell, Gasc\'{o}n, and Nissim \cite{BBGN19-3} for \emph{statistically secure} aggregation has a message complexity that only depends logarithmically on those two parameters.

If it is not possible to improve our shuffler-based implementation of the aggregator, it might still be possible to improve alternative implementations. For example, the seminal protocol by Ben-Or, Goldwasser, and Wigderson \cite{BGW88} offers perfect security when there is an honest majority; could we tweak the construction to improve communication guarantees while satisfying the weaker guarantee of LLR security? More ambitiously, one could imagine hardness assumptions being stated in terms of LLR distance instead of statistical distance. In that case, cryptographic tools would facilitate pure differential privacy against bounded adversaries.

\section*{Acknowledgements}
The authors are members of the Data Co-Ops project (https://datacoopslab.org). This work was supported in part by a gift to Georgetown University. We would like to thank Kobbi Nissim for his assistance regarding the definitions and terminology, as well as correcting an oversight in our impossibility result. Gautam Kamath and Cl{\'{e}}ment Canonne gave helpful comments regarding related work in private uniformity testing.

\bibliographystyle{plain}
\bibliography{refs}

\appendix

\section{Deferred Proofs for Section \ref{sec:sim-mod}}
Here, we present the proofs of the technical steps required for our security argument. Recall that, in Section \ref{sec:sim-mod}, we use $\bD$ as shorthand for the truncated discrete Laplace distribution $\DLap_{t/2}(t/2,\exp(-\lambda))$.

\subsection{Analysis of $\alpha_z,\alpha'_z$}
\label{sec:alpha}

\begin{proof}[Proof of Claim \ref{clm:easy-case}]
We begin with the case where $z=x_1+x_2+vm$ for some integer $v\in[0,2t]$. We can expand the term $\alpha_z$ as
\begin{align*}
\alpha_z &= \sum_{u=0}^t \bD_{x_1}[x_1+um]\cdot \bD_{x_2}[z-(x_1+um)]\\
    &= \sum_{u=0}^t \bD_{x_1}[x_1+um]\cdot \bD_{x_2}[x_2+(v-u)m] \tag{Structure of $z$}\\
    &= \sum_{u=0}^t \bD[u]\cdot \bD[v-u]
\end{align*}

Because $x_1+x_2 = x'_1 + x'_2$, we also have that $z=x'_1+x'_2+vm$ so completely symmetric steps yield the equality $\alpha'_z = \sum_{u=0}^t \bD[u]\cdot \bD[v-u] =\alpha_z$.

In the case where no integer $v\in[0,2t]$ exists, observe that this implies neither $(\bD_{x_1} * \bD_{x_2})$ nor $(\bD_{x'_1} * \bD_{x'_2})$ can produce $z$. So $\alpha'_z=0=\alpha_z$.
\end{proof}

\begin{proof}[Proof of Claim \ref{clm:bounding-alpha}]
If no $v$ exists, then it is not possible for either $\bD_{x_1}*\bD_{x_2}$ or $\bD_{x'_1}*\bD_{x'_2}$ to produce $z$ so $\alpha_z = 0 = \alpha'_z$.

Otherwise, $z$ must be represented in Table \ref{tab:alpha}. Notice that one of two cases must hold: either $\alpha_z$ has one more non-zero term than $\alpha'_z$ or vice versa. We argue (a) this term takes the form $\bD[\min(v,2t-v)] \cdot \bD[0]$ and (b) the other terms in the two sums can be paired off such that each pair's members are within a factor of $e^\lambda$ of one another.

\begin{table}[ht]
    \centering
    \begin{tabular}{|l|c|c|}
        \hline
         $z$ & $\alpha_z$ & $\alpha'_z$  \\ \hline \hline
         $x_1 + x_2$ & $\bD_{x_1}[x_1]\cdot \bD_{x_2}[x_2]$ & 0\\ \hline
         $x_1 + x_2 + m$ & $\bD_{x_1}[x_1]\cdot \bD_{x_2}[x_2 + m]$ & $\bD_{x'_1}[x'_1]\cdot \bD_{x'_2}[x'_2]$ \\ 
         $=x'_1+x'_2$ & $+ \bD_{x_1}[x_1 + m]\cdot \bD_{x_2}[x_2]$ & \\ \hline
         \dots & \dots & \dots \\ \hline
         $x_1 + x_2 + 2tm $ & $\bD_{x_1}[x_1+tm] \cdot \bD_{x_2}[x_2+ tm]$ & $\bD_{x'_1}[x'_1 + (t-1)m]\cdot \bD_{x'_2}[x'_2 + tm]$ \\
         $=x'_1+x'_2 + (2t-1)m$ & & $+ \bD_{x'_1}[x'_1 + tm]\cdot \bD_{x'_2}[x'_2 + (t-1)m]$ \\ \hline
         $x_1 + x_2 + (2t+1)m $ & 0 & $\bD_{x'_1}[x'_1+tm]\cdot \bD_{x'_2}[x'_2+tm]$ \\
         $=x'_1+x'_2 + 2tm$ & & \\ \hline
    \end{tabular}
    \caption{Select choices of $z$ and corresponding $\alpha_z,\alpha'_z$ values.}
    \label{tab:alpha}
\end{table}

For $v\in [0,t]$,
\begin{align*}
\alpha_z &= \sum^v_{u=0} \bD_{x_1}[x_1+um]\cdot \bD_{x_2}[x_2 + (v-u)m]\\
    &= \sum^v_{u=0} \bD[u]\cdot \bD[v-u]
\end{align*}
where each term in the summation is positive. Meanwhile,
\begin{align*}
\alpha'_z &= \sum^{v-1}_{u=0} \bD_{x'_1}[x'_1+um]\cdot \bD_{x'_2}[x'_2 + (v-1-u)m]\\
    &= \sum^{v-1}_{u=0} \bD[u]\cdot \bD[v-1-u].
\end{align*}
such that, for $v>0$, each term is positive. Due to the definition of $D$, note that $\bD[v-u]$ and $\bD[v-1-u]$ are within an $e^\lambda$ factor of one another. And the last term in $\alpha_z$ is $\bD[v]\cdot \bD[0]$.

In the case where $v \in [t+1,2t+1]$, we can simply use the symmetry of $D$ and the symmetry of Table \ref{tab:alpha} to observe
\begin{align*}
\alpha_z &= \sum_{u=0}^{2t-v} \bD[u]\cdot \bD[(2t-v)-u]\\
\alpha'_z &= \sum_{u=0}^{2t+1-v} \bD[u]\cdot \bD[(2t+1-v)-u]
\end{align*}
As before, $\bD[(2t-v)-u]$ and $\bD[(2t+1-v)-u]$ are within an $e^\lambda$ factor of one another. And the last term in $\alpha'_z$ is $\bD[2t+1-v]\cdot \bD[0]$.
\end{proof}

\begin{proof}[Proof of Claim \ref{clm:extra-alpha-term-1}]
\underline{Case 1}: $v\leq t$: $\bD[\min(v,2t+1-v)] = \bD[v]$. Now, the ratio of interest is
\begin{align*}
\frac{\bD[v] \cdot \bD[0]}{\alpha'_z} &= \frac{\bD[v] \cdot \bD[0]}{\sum^{v-1}_{u=0} \bD[u]\cdot \bD[v-1-u]} \\
    &\leq \frac{\bD[v] \cdot \bD[0]}{\sum^{v-1}_{u=0} \bD[u]\cdot \bD[0]} \tag{Shape of $D$}\\
    &= \frac{\bD[v]}{\sum^{v-1}_{u=0} \bD[u]} \\
    &= \frac{\exp(-\lambda|\tfrac{t}{2}-v|)}{\sum_{u=0}^{v-1} \exp(-\lambda|\tfrac{t}{2}-u|)} \stepcounter{equation} \tag{\theequation} \label{eq:extra-alpha-term-1}
\end{align*}
We split into more cases.

\underline{Case 1a}: $v < t/2$. The denominator is a geometric series with common ratio $r = \exp(-\lambda)$, coefficient $a = \exp(-\lambda \cdot |\tfrac{t}{2}-v+1|)$, and $v$ terms. So we have $$a\sum_{u=0}^{v-1} r^u = a\cdot \left( \frac{1-r^v}{1-r} \right) = \exp(-\lambda \cdot |\tfrac{t}{2}-v+1|) \cdot \left( \frac{1-\exp(-\lambda v)}{1-\exp(-\lambda)} \right),$$
which in turn means
\begin{align*}
\eqref{eq:extra-alpha-term-1} &=  e^\lambda \cdot \frac{1-e^{-\lambda}}{1-e^{-\lambda v}}  \\
&= \frac{e^\lambda-1}{1-e^{-\lambda v}} \\
&\leq \frac{e^\lambda-1}{1-1/e} \tag{$v\geq 1/\lambda$} \\
&\leq 3\lambda
\end{align*}
The last inequality follows from $\lambda < 1$.

\underline{Case 1b}: $v \in [t/2, t-1/\lambda]$. We lean on the symmetry of $D$ (meaning, $\bD[v]=\bD[t-v]$):
\begin{align*}
\eqref{eq:extra-alpha-term-1} &=\frac{\exp(-\lambda|\tfrac{t}{2}-(t-v)|)}{\sum_{u=0}^{v-1} \exp(-\lambda|\tfrac{t}{2}-u|)} \\
&\leq \frac{\exp(-\lambda|\tfrac{t}{2}-(t-v)|)}{\sum_{u=0}^{(t-v)-1} \exp(-\lambda|\tfrac{t}{2}-u|)}
\end{align*}
so we can repeat the same argument as before, this time using $t-v$ instead of $v$ and $v \leq t-1/\lambda$ instead of $v \geq 1/\lambda$. 

\underline{Case 1c}: $v \in (t-1/\lambda, t]$. We combine the symmetry of $D$ along with the fact that the left side is monotonically increasing:
\begin{align*}
\eqref{eq:extra-alpha-term-1} &\leq \frac{\exp(-\lambda|\tfrac{t}{2}-1/\lambda|)}{\sum_{u=0}^{v-1} \exp(-\lambda|\tfrac{t}{2}-u|)} \\
&\leq \frac{\exp(-\lambda|\tfrac{t}{2}-1/\lambda|)}{\sum_{u=0}^{1/\lambda-1} \exp(-\lambda|\tfrac{t}{2}-u|)} 
\end{align*}
We can now repeat the argument from Case 1a.

\underline{Case 2}: In the case where $v > t$, $\bD[\min(v,2t+1-v)] = \bD[2t+1-v]$. Define the shorthand $\overline{v} := 2t+1-v$ Observe that $\overline{v} \leq t$. We can therefore repeat the argument from the $v\leq t$ case, simply using $\overline{v}$ in place of $v$.
\end{proof}

\begin{proof}[Proof of Claim \ref{clm:extra-alpha-term-2}]
Consider any $v\neq 2t+1$.
\begin{align*}
\delta_z &= (B*B)[x_1+x_2+vm] \\
    &\geq \bB[x_1 + \min(v,2t+1-v)\cdot m]\cdot \bB[x_2 + (v- \min(v,2t+1-v)) \cdot m] \\
    &= \frac{1}{m^2} \cdot \bL[\min(v,2t+1-v)]\cdot \bL[v-\min(v,2t+1-v)] \stepcounter{equation} \label{eq:extra-alpha-term-2,1} \tag{\theequation} \\
    &\geq \frac{1}{m^2} \cdot \bL[\min(v,2t+1-v)]\cdot \bD[0] \tag{Defn. of $D,L$} \\
    &= \frac{1}{m^2} \cdot \frac{\bL[\min(v,2t+1-v)]}{\bD[\min(v,2t+1-v)]} \cdot \bD[\min(v,2t+1-v)]\cdot \bD[0] \stepcounter{equation} \label{eq:extra-alpha-term-2,2} \tag{\theequation}
\end{align*}
We briefly justify \eqref{eq:extra-alpha-term-2,1}. The constraint $v \neq 2t+1$ implies both $\min(v,2t+1-v)$ and $v-\min(v,2t+1-v)$ are at most $t$. In turn, both $B$ terms can be equated with corresponding $L$ terms.

For $v=2t+1$, we arrive at the same expression but this time using the equality $\delta_z =  (B*B)[x'_1+x'_2+v'\cdot m]$, for $v' = v-1 =2t$ so we can equate $B$ terms with $L$ terms.

Now we expand the ratio inside \eqref{eq:extra-alpha-term-2,2}. For $v \leq 1/\lambda$, we have that $v < 2t+1-v$ so
\begin{align*}
\frac{\bL[\min(v,2t+1-v)]}{\bD[\min(v,2t+1-v)]} &= \frac{\bL[v]}{\bD[v]} \\
&= \frac{\bD[\lfloor t/2-v \rfloor]}{\bD[v]} \tag{$v< 1/\lambda < t/2$} \\
&= \frac{\exp(-\lambda\cdot |1/2 +v|)}{\exp(-\lambda \cdot |t/2 -v|)}\\
&= \exp(\lambda \cdot (t/2-2v-1/2)) \\
&\geq \exp(\lambda \cdot (t/2-2/\lambda-1/2)) \tag{Bound on $v$}\\
&\geq \frac{m^2}{\lambda\cdot p^2} \tag{Bound on $t$}
\end{align*}

For $v \geq 2t+1-1/\lambda$, we have that $v > 2t+1-v$ so
\begin{align*}
\frac{\bL[\min(v,2t+1-v)]}{\bD[\min(v,2t+1-v)]} &= \frac{\bL[2t+1-v]}{\bD[2t+1-v]} \\
    &= \frac{\bD[\lfloor t/2 \rfloor -(2t+1-v)]}{\bD[2t+1-v]} \tag{$2t+1-v <t/2$} \\
    &= \frac{\exp(-\lambda\cdot (1/2+ (2t+1-v)))}{\exp(-\lambda\cdot (-3t/2 - 1 +v ))} \\
    &= \exp(\lambda\cdot (-7t/2 + 2v -5/2))\\
    &\geq \exp(\lambda\cdot (t/2 -2/\lambda -1/2) \geq \frac{m^2}{\lambda\cdot p^2}
\end{align*}
The claim follows by substitution into \eqref{eq:extra-alpha-term-2,2}
\end{proof}

\subsection{Analysis of $\beta_z,\beta'_z$}
\label{sec:beta}

\begin{proof}[Proof of Claim \ref{clm:bounding-beta}]
If $z<\min(x,x')$, then both $(D_x*B)[z]$ and $(D_x*B)[z]$ are zero so the claim trivially holds. Otherwise, let $c:= \lfloor (z-x) / m \rfloor$ and $c':= \lfloor (z-x') / m \rfloor$, then $u = \min(t, c)$ and $u' = \min(t, c')$.
\begin{align*}
(\bD_x*\bB)[z] &= \sum_{w =0}^u \bD_{x}[x+wm] \cdot \bB[z - (x+wm)]\\
    &= \sum_{w =0}^u \bD[w] \cdot \bB[z - (x+wm)]
\end{align*}
\begin{align*}
(\bD_{x'}*\bB)[z] &= \sum_{w=0}^{u'} \bD_{x'}[x'+wm] \cdot \bB[z - (x'+wm)]\\
    &= \sum_{w=0}^{u'} \bD[w] \cdot \bB[z - (x'+wm)]
\end{align*}

Because $|x-x'|< m$, notice that $|c-c'|\leq 1$ which in turn means $|u-u'|\leq 1$. We can pair off all terms in the two summations except for at most one. If $u=u'+1$, then the extra term is $\bD[u]\cdot \bB[z-(x+um)]$. Otherwise, it is $\bD[u']\cdot \bB[z-(x'+um)]$. Without loss of generality, we focus on the first case:
\begin{align*}
(\bD_x*\bB)[z] &= \sum_{w =0}^{u-1} \bD[w] \cdot \bB[z - (x+wm)] + \bD[u]\cdot \bB[z-(x+um)]\\
    &\leq e^\lambda \cdot \sum_{w =0}^{u-1} \bD[w] \cdot \bB[z - (x'+wm)] + \bD[u]\cdot \bB[z-(x+um)] \stepcounter{equation} \label{eq:bounding-beta-1} \tag{\theequation}
\end{align*}

We further split into cases regarding $u$.

\underline{Case 1}: $u \notin [1/\lambda, t-1/\lambda]$. We will show $\bD[u] \leq \lambda\cdot p \cdot \bB[x+um]$ so that
\begin{align*}
\eqref{eq:bounding-beta-1} &\leq e^\lambda \cdot \sum_{w =0}^{u-1} \bD[w] \cdot \bB[z - (x'+wm)] + \lambda\cdot p\cdot \bB[x+um]\cdot \bB[z-(x+um)] \\
    &\leq e^\lambda \cdot \sum_{w =0}^{u-1} \bD[w] \cdot \bB[z - (x'+wm)] + \lambda\cdot p \cdot (\bB*\bB)[z] \\
    &\leq e^\lambda \cdot (\bD_{x'}*B)[z] + \lambda\cdot p \cdot (\bB*\bB)[z]
\end{align*}
which would complete our argument. We begin with the $u<1/\lambda$ case:
\begin{align*}
\frac{\bD[u]}{\bB[x+um]} &= m\cdot \frac{\bD[u]}{\bL[u]} \\
    &= m\cdot \frac{\bD[u] }{\bD[\lfloor t/2\rfloor - u]} \tag{Defn. of $L$}\\
    &= m\cdot \frac{\exp(-\lambda\cdot (t/2-u) ) }{\exp(-\lambda\cdot (u+1/2))} \\
    &= m\cdot \exp(\lambda\cdot (-t/2 + 2u + 1/2))\\
    &\leq m\cdot \exp(\lambda\cdot (-t/2 + 2/\lambda + 1/2)) \tag{$u\leq 1/\lambda$}\\
    &\leq \lambda\cdot p 
\end{align*}
The second and last steps follow from our lower bound on $t$. For $u > t-1/\lambda$, we repeat the same steps
\begin{align*}
\frac{\bD[u]}{\bB[x+um]} &= m\cdot \frac{\bD[u]}{\bL[u]} \\
    &= m\cdot \frac{\bD[u] }{\bD[u-\lceil t/2\rceil]} \tag{Defn. of $L$}\\
    &= m\cdot \frac{\exp(-\lambda\cdot (u-t/2)}{\exp(-\lambda\cdot (t-u+1/2)}\\
    &= m\cdot \exp(\lambda\cdot(3t/2-2u+1/2))\\
    &< m\cdot \exp(\lambda\cdot(-t/2+2/\lambda+1/2)) \tag{$u > t-1/\lambda$}\\
    &\leq \lambda\cdot p
\end{align*}

\underline{Case 2}: $u \in [1/\lambda, t-1/\lambda]$,
\begin{align*}
\frac{\bD[u]\cdot \bB[z-(x+um)]}{\sum_{w=0}^{u-1} \bD[w]\cdot \bB[z-(x'+wm)]} &= \frac{\bD[u]\cdot \bL[\lfloor (z-x-um) / m \rfloor ]}{\sum_{w=0}^{u-1} \bD[w]\cdot \bL[\lfloor (z-x'-wm) / m \rfloor]}\\
&= \frac{\bD[u]\cdot \bL[\lfloor (z-x) / m \rfloor -u ]}{\sum_{w=0}^{u-1} \bD[w]\cdot \bL[\lfloor (z-x') / m \rfloor - w]} \\
&= \frac{\bD[\min(t,c)]\cdot \bL[c -\min(t,c) ]}{\sum_{w=0}^{\min(t,c)-1} \bD[w]\cdot \bL[c' - w]}\\
&= \frac{\bD[c]\cdot \bL[0]}{\sum_{w=0}^{c-1} \bD[w]\cdot \bL[c-1-w]} \stepcounter{equation} \label{eq:bounding-beta-2} \tag{\theequation}
\end{align*}
The final step follows from $c=\min(t,c)$, which itself comes from $u = \min(t,c) \in [1/\lambda, t-1/\lambda]$ and the lower bound on $t$.

\underline{Case 2a}: $c \leq \lfloor t/2\rfloor$. By definition of $L$,
\begin{align*}
\eqref{eq:bounding-beta-2} &= \frac{\bD[c]\cdot \bD[\lfloor t/2 \rfloor ]}{\sum_{w=0}^{c-1} \bD[w]\cdot \bD[\lfloor t/2 \rfloor - (c-1-w)]}\\
    &= \frac{\exp(-\lambda \cdot |t/2 - c| ) \cdot \exp(-\lambda/2)}{\sum_{w=0}^{c-1} \exp(-\lambda \cdot |t/2 - w| ) \cdot \exp(-\lambda\cdot |1/2 +(c-1-w)|)}\\
    &= \frac{\exp(-\lambda \cdot (t/2 - c + 1/2) ) }{\sum_{w=0}^{c-1} \exp(-\lambda \cdot (t/2 +c - 2w - 1/2) ) } \tag{$c \leq \lfloor t/2\rfloor$}\\
    &= \frac{1}{\sum_{w=0}^{c-1} \exp(\lambda\cdot (t/2-c+1/2 -(t/2+c-2w-1/2)))}\\
    &= \frac{1}{\sum_{w=0}^{c-1} \exp(\lambda\cdot (1-2c+2w))} \stepcounter{equation} \label{eq:bounding-beta-3} \tag{\theequation}
\end{align*}

We once again must evaluate a geometric series. This time, we have common ratio $r=e^{-2\lambda}$, coefficient $a=\exp(\lambda\cdot(1-2c+2(c-1))) = e^{-\lambda}$ and $c$ terms:
$$a\sum_{w=0}^{c-1} r^w = a\cdot \left( \frac{1-r^c}{1-r} \right) = e^{-\lambda}\cdot \frac{1-e^{-2c\lambda}}{1-e^{-2\lambda}}$$
Therefore, 
\begin{align*}
\eqref{eq:bounding-beta-3} &= e^\lambda \cdot \frac{1-e^{-2\lambda}}{1-e^{-2c\lambda}}\\
    &\leq \frac{e^\lambda-e^{-\lambda}}{1-e^{-2}} \tag{$c > 1/\lambda$}\\
    &\leq 3\lambda \tag{$\lambda \leq 1$}
\end{align*}

\underline{Case 2b}: $c>\lfloor t/2\rfloor$. Let $\kappa$ be shorthand for $\lceil 1/\lambda \rceil$. By definition of $L$,
\begin{align*}
\eqref{eq:bounding-beta-2} &= \frac{\bD[c]\cdot \bD[\lfloor t/2\rfloor]}{\sum_{w=0}^{c-1} \bD[w]\cdot \bL[c-1-w]} \\
    &\leq \frac{\bD[c]\cdot \bD[\lfloor t/2 \rfloor]}{\sum_{w=\lfloor t/2 \rfloor-\kappa+1}^{\lfloor t/2 \rfloor} \bD[w]\cdot \bL[c-1-w] } \stepcounter{equation} \label{eq:bounding-beta-4} \tag{\theequation}
\end{align*}
We unpack the summation in the denominator. When $w=\lfloor t/2\rfloor$, the corresponding term is
\begin{align*}
&\bD[\lfloor t/2\rfloor]\cdot \bL[c-1-\lfloor t/2\rfloor]\\
={}& \bD[\lfloor t/2\rfloor]\cdot \bD[\lfloor t/2\rfloor - (c-1-\lfloor t/2\rfloor)] \tag{Defn. of $L$}\\
={}& \bD[\lfloor t/2\rfloor]\cdot \bD[(t-1)-(c-1)] \tag{$t$ odd}\\
={}& \bD[\lfloor t/2\rfloor]\cdot \bD[c] \tag{Symmetry of $D$}
\end{align*}
For any smaller $w$, notice that $\bD[w]\cdot \bL[c-1-w]$ and $\bD[w+1]\cdot \bL[c-1-(w+1)]$ are within an $e^{2\lambda}$ multiplicative factor of one another by construction. Thus, we can lower bound the summation by a geometric series with common ratio $r=e^{-2\lambda}$, coefficient $a = \bD[\lfloor t/2\rfloor]\cdot \bD[c]$, and $\kappa$ terms:
$$ a \cdot \sum_{w=0}^{\kappa-1} r^w = a \cdot \frac{1-r^\kappa}{1-r}  = \bD[\lfloor t/2\rfloor]\cdot \bD[c]\cdot \frac{1-e^{-2\kappa\lambda}}{1-e^{-2\lambda}} $$
By substitution, we have
\begin{align*}
\eqref{eq:bounding-beta-4} \leq \frac{\bD[\lfloor t/2\rfloor]\cdot \bD[c]}{\bD[\lfloor t/2\rfloor]\cdot \bD[c] \cdot \frac{1-e^{-2\kappa\lambda}}{1-e^{-2\lambda}}} &= \frac{1-e^{-2\lambda}}{1-e^{-2\kappa\lambda}}\\
    &\leq \frac{1-e^{-2\lambda}}{1-e^{-2}} \tag{$\kappa \geq 1/\lambda$}\\
    &\leq 3\lambda
\end{align*}
This concludes all cases of the proof.
\end{proof}

\subsection{Lower Bound for Perfectly Secure Aggregation}
\label{apdx:sim-perfect-mod}

\begin{clm*}[Copy of Claim \ref{clm:perfect-lower-bound}]
Fix any number of users $n \geq 2$, message complexity $k \in \N$, and modulus $m \geq 2$. If $(S^{n,2,k} \circ \vec{R})$ is an $(0,q)$-relaxation of $\Sigma^{n,m,1}$, then $q > 1-\frac{1}{m}$.
\end{clm*}

\begin{proof}
At a high level, we prove that perfect security implies that the behavior of randomizer $R_2$ on input $0$ is identical to its behavior on $m-1$. Our argument is without loss of generality, so that the behavior of the construction $(S^{n,2,k} \circ \vec{R})$ on any input is the same as any other. Thus, there is no \POST~ function which can do better than guessing uniformly at random from $\{0,\dots,m-1\}$.

Consider the central model algorithm $C$ which, on input $\vec{x}$, computes the sum (without modulus) of all $nk$ bits in $(S^{2,k} \circ \vec{R})(\vec{x})$. Using ``$=_{\textrm{dist}}$'' to denote equality in distribution, perfect security implies
\begin{align*}
C(\underbrace{0,\dots,0}_{n}) &=_{\textrm{dist}} C(1,m-1,\underbrace{0,\dots,0}_{n-2}) \\
C(1,\underbrace{0,\dots,0}_{n-1}) &=_{\textrm{dist}} C(2,m-1,\underbrace{0,\dots,0}_{n-2}) \\
C(2,\underbrace{0,\dots,0}_{n-1}) &=_{\textrm{dist}} C(3,m-1,\underbrace{0,\dots,0}_{n-2}) \\
\dots \\
C(m-2,\underbrace{0,\dots,0}_{n-1}) &=_{\textrm{dist}} C(m-1,m-1,\underbrace{0,\dots,0}_{n-2})
\end{align*}

For any randomizer $R_i$ and value $x\in \{0,\dots,m-1\}$, let $\hat{R}_i(x)$ be the distribution of $\sum_{j\in[k]} y_{i,j}$ (without modulus), where $y_{i,1},\dots,y_{i,k}$ are the binary messages produced by $R_i(x)$. Now consider the algorithm $C'$ which, on input $\vec{x}$, samples $y_{i,1},\dots,y_{i,k} \sim \hat{R}_i(x_i)$ for each $i\in [n]$ and then reports $\sum_{i\in n} \sum_{j\in[k]} y_{i,j}$ (without modulus). Because our shuffler is perfectly correct and addition is commutative,
$$
\forall \vec{x} ~ C(\vec{x}) =_{\textrm{dist}} C'(\vec{x})
$$ 
By construction, we also have
$$
C'(\vec{x}) =_{\textrm{dist}} \hat{R_1}(x_1) *\dots * \hat{R_n}(x_n),
$$
where we use $A*B$ to denote the convolution between distributions $A,B$.

Taken together, the above implies 
\begin{align}
\hat{R_1}(0) * \underbrace{\hat{R_2}(0) * \dots * \hat{R_n}(0)}_{n-1~\textrm{randomizers with input 0}} &=_{\textrm{dist}} \hat{R_1}(1) * \hat{R_2}(m-1) * \underbrace{\hat{R_3}(0) * \dots * \hat{R_n}(0)}_{n-2~\textrm{randomizers with input 0}} \label{eq:impossibility-1} \\
\hat{R_1}(1) * \underbrace{\hat{R_2}(0) * \dots * \hat{R_n}(0)}_{n-1~\textrm{randomizers with input 0}} &=_{\textrm{dist}} \hat{R_1}(2) * \hat{R_2}(m-1) * \underbrace{\hat{R_3}(0) * \dots * \hat{R_n}(0)}_{n-2~\textrm{randomizers with input 0}} \\
\hat{R_1}(2) * \underbrace{\hat{R_2}(0) * \dots * \hat{R_n}(0)}_{n-1~\textrm{randomizers with input 0}} &=_{\textrm{dist}} \hat{R_1}(3) * \hat{R_2}(m-1) * \underbrace{\hat{R_3}(0) * \dots * \hat{R_n}(0)}_{n-2~\textrm{randomizers with input 0}} \\
&\dots \nonumber \\
\hat{R_1}(m-2) * \underbrace{\hat{R_2}(0) * \dots * \hat{R_n}(0)}_{n-1~\textrm{randomizers with input 0}} &=_{\textrm{dist}} \hat{R_1}(m-1) * \hat{R_2}(m-1) * \underbrace{\hat{R_3}(0) * \dots * \hat{R_n}(0)}_{n-2~\textrm{randomizers with input 0}} \\
\hat{R_1}(m-1) * \underbrace{\hat{R_2}(0) * \dots * \hat{R_n}(0)}_{n-1~\textrm{randomizers with input 0}} &=_{\textrm{dist}} \hat{R_1}(0) * \hat{R_2}(m-1) * \underbrace{\hat{R_3}(0) * \dots * \hat{R_n}(0)}_{n-2~\textrm{randomizers with input 0}} \label{eq:impossibility-2}
\end{align}


Let $\mu_{i,x}(v)$ be the moment generating function (MGF) of $\hat{R_i}(x)$.\footnote{This MGF exists because $\hat{R_i}(x)$ is a distribution over the finite integers $\{0,\dots,k\}$.} Recall that if $a,b$ are independent random variables (drawn from $A,B$) with MGFs $\mu_a(v), \mu_b(v)$, then the MGF of $a+b$ (drawn from $A*B$) is $\mu_a(v)\cdot \mu_b(v)$. Because \eqref{eq:impossibility-1} to \eqref{eq:impossibility-2} are equalities in distribution, it must be the case that
\begin{align*}
\mu_{1,0}(v) \cdot \mu_{2,0}(v) \cdot \prod_{i=3}^n{\mu_{i,0}(v)} &= \mu_{1,1}(v) \cdot \mu_{2,m-1}(v) \cdot \prod_{i=3}^n{\mu_{i,0}(v)} \\
\mu_{1,1}(v) \cdot \mu_{2,0}(v) \cdot \prod_{i=3}^n{\mu_{i,0}(v)} &= \mu_{1,2}(v) \cdot \mu_{2,m-1}(v) \cdot \prod_{i=3}^n{\mu_{i,0}(v)} \\
\mu_{1,2}(v) \cdot \mu_{2,0}(v) \cdot \prod_{i=3}^n{\mu_{i,0}(v)} &= \mu_{1,3}(v) \cdot \mu_{2,m-1}(v) \cdot \prod_{i=3}^n{\mu_{i,0}(v)} \\
\dots& \\
\mu_{1,m-2}(v) \cdot \mu_{2,0}(v) \cdot \prod_{i=3}^n{\mu_{i,0}(v)} &= \mu_{1,m-1}(v) \cdot \mu_{2,m-1}(v) \cdot \prod_{i=3}^n{\mu_{i,0}(v)}\\
\mu_{1,m-1}(v) \cdot \mu_{2,0}(v) \cdot \prod_{i=3}^n{\mu_{i,0}(v)} &= \mu_{1,0}(v) \cdot \mu_{2,m-1}(v) \cdot \prod_{i=3}^n{\mu_{i,0}(v)}
\end{align*}
If we take the product of all these equalities, we have
$$
\left( \prod_{x=0}^{m-1} \mu_{1,x}(v) \right) \cdot \mu_{2,0}^{m}(v) \cdot \left( \prod_{i=3}^n \mu_{i,0}^m(v) \right) = \left( \prod_{x=0}^{m-1} \mu_{1,x}(v) \right) \cdot \mu_{2,m-1}^{m}(v) \cdot \left( \prod_{i=3}^n \mu_{i,0}^m(v) \right)
$$
Because moment generating functions are positive, we can cancel terms on both sides of the above and conclude that $\mu_{2,0}(v) = \mu_{2,m-1}(v)$. And when the moment generating functions of two random variables are the same at every point, the random variables are identically distributed: $\hat{R}_2(0)$ and $\hat{R}_2(m-1)$ are identical.

Observe that our construction was without loss of generality: we have, for all $i,x,x'$, $\hat{R}_i(x) =_\textrm{dist} \hat{R}_i(x')$. Consider the algorithm $M_i$ which, on input $x$, computes $y\gets \hat{R}_i(x)$ and then generates a binary vector where the prefix is $y$ ones. Naturally, we have that 
\begin{equation}
\label{eq:impossibility-3}
M_i(x)=_\textrm{dist} M_i(x')
\end{equation}

Although the distribution of $M_i(x)$ may not be the same as $R_i(x)$, notice that the total mass placed by $R_i(x)$ on binary vectors with sum $y$ is the same as the mass placed by $M_i(x)$ on the binary vector where the prefix is $y$ ones. Therefore, by the definition of our shuffler, we have that
\begin{equation}
\label{eq:impossibility-4}
(S^{n,2,k} \circ \vec{R})(\vec{x}) =_\textrm{dist} (S^{n,2,k} \circ \vec{M})(\vec{x})
\end{equation}
for any input $\vec{x}$. We use this to prove $(S^{n,2,k} \circ \vec{R})(\vec{x}) =_\textrm{dist} (S^{n,2,k} \circ \vec{R})(\vec{x}\,')$ for any other input $\vec{x}\,'$:
\begin{align*}
(S^{n,2,k} \circ \vec{R})(\vec{x}) &=_\textrm{dist} (S^{n,2,k} \circ \vec{M})(\vec{x}) \tag{From \eqref{eq:impossibility-4}} \\
    &=_\textrm{dist} (S^{n,2,k} \circ \vec{M})(\vec{x}\,') \tag{From \eqref{eq:impossibility-3}} \\
    &=_\textrm{dist} (S^{n,2,k} \circ \vec{R})(\vec{x}\,') \tag{From \eqref{eq:impossibility-4}}
\end{align*}
This concludes the proof, since no $\POST$ algorithm can recover information about the input.
\end{proof}





\begin{clm*}[Copy of Claim \ref{clm:binary-reduction}]
For any shuffle protocol $P=(R,S^{m,k},A)$, there is a shuffle protocol that uses binary messages $P'=(R',S^{2,k'},A')$ and exactly simulates $P$: for any input $\vec{x}$, $P'(\vec{x})$ is identically distributed with $P(\vec{x})$.
\end{clm*}

\begin{proof}
By the definition of the shuffler, the output of $(S^{m,k}\circ \vec{R})(\vec{x})$ can be exactly simulated from the function which computes the histogram of messages generated by $\vec{R}(\vec{x})$. Notice that an $m$-bin histogram of $nk$ values can be represented as an integer between 0 and $(nk)^m$. More to the point, each message logged by the histogram can be encoded as an integer between 0 and $(nk)^m$ such that the sum of all such encodings is the histogram.

Let $T^{m,k}$ be the algorithm that takes as input a vector $\vec{w} \in \{0,\dots,m-1\}^k$ and computes the integer $t_{\vec{w}}$ such that the $j$-th digit of $t_{\vec{w}}$ in base-$nk$ is $w_j$.

Let $k'\gets (nk)^m$. Let $R'_i$ be the algorithm that, on input $x$, computes $u\gets T^{m,k}(R_i(x))$ and outputs a vector of $u$ ones and $k'-u$ zeroes. Let $A'$ be the algorithm that, on input $\vec{y}$, computes $t\gets \sum y_i$, constructs $\vec{w}\in \{0,\dots,m-1\}^{nk}$ such that the frequency of $j$ in $\vec{w}$ is the $j$-th digit of $t$ in base-$nk$, and reports $A(\vec{w})$.

The transformation from messages to integers and back again is lossless, so the simulation is exact.
\end{proof}


\section{Proofs for Uniformity Testing Protocols}
\label{apdx:ut}
Before we dive into our proofs, we first build an understanding of the truncated discrete Laplace distribution. Specifically, we bound the second and fourth moments of $\DLap_\tau(\lambda)$ when $\tau$ is sufficiently large.
\begin{clm}
\label{clm:DLap}
Fix any $\tau\in \N$ and $\lambda \in (0,1)$ such that $\tau > \frac{\ln 10}{\ln (1/\lambda)} - 1$. The first, second, and fourth moments of $\DLap_\tau(\lambda)$ are 0, $< \frac{5}{(1-\lambda)^2}$, and $< \frac{60}{(1-\lambda)^4}$, respectively.
\end{clm}
\begin{proof}
The fact that the expectation of $\DLap_\tau(\lambda)$ is 0 is immediate from the mass function of $\DLap(\lambda)$ and the symmetry of the truncation definition. So we devote the rest of the proof to the second and fourth moments: for either $k\in\{2,4\}$,
\begin{align*}
\ex{\eta \sim \DLap_\tau(\lambda)}{\eta^k} &= \sum_{i\neq 0} i^k \cdot \pr{\eta \sim \DLap_\tau(\lambda)}{\eta = i}\\
    &= 2\cdot \sum_{i=1}^\tau i^k\cdot \pr{\eta \sim \DLap_\tau(\lambda)}{\eta = i} \tag{Symmetry} \\
    &= 2\cdot \sum_{i=1}^\tau i^k\cdot \frac{\lambda^i}{1+ 2\sum_{j=1}^\tau \lambda^j} \tag{By definition}\\
    &= \frac{2}{1+ 2(\frac{1-\lambda^{\tau+1}}{1-\lambda}-1)} \cdot \sum_{i=1}^\tau i^k\cdot \lambda^i \\
    &= \frac{2}{1+\lambda-2\lambda^{\tau+1}} \cdot \sum_{i=1}^\tau i^k\cdot \lambda^i \cdot (1-\lambda) \stepcounter{equation} \label{eq:DLap-moments-1} \tag{\theequation}
\end{align*}

We focus our attention on the leading ratio. Specifically, we derive a bound on the term $2\lambda^{\tau+1}$ from our bound on $\tau$:
\begin{align*}
    \tau &> \frac{\ln 10}{\ln (1/\lambda)} - 1\\
    (\tau+1) \ln (1/\lambda) &> \ln 10\\
    (\tau+1) \ln \lambda &< \ln (1/10)\\
    2\lambda^{\tau+1} &< 1/5
\end{align*}

Thus,
\begin{align*}
\eqref{eq:DLap-moments-1} &< \frac{2}{(4/5)+\lambda} \cdot \sum_{i=1}^\tau i^k\cdot \lambda^i \cdot (1-\lambda) \\
    &< \frac{5}{2} \cdot \sum_{i=1}^\tau i^k\cdot \lambda^i \cdot (1-\lambda) \tag{$\lambda >0 $}\\
    &< \frac{5}{2} \cdot \sum_{i=0}^\infty i^k\cdot \lambda^i \cdot (1-\lambda)\\
    &= \frac{5}{2} \cdot \ex{\eta \sim \Geo(1-\lambda)}{\eta^k} \stepcounter{equation} \label{eq:DLap-moments-2} \tag{\theequation}
\end{align*}
where $\Geo(p)$ denotes the (geometric) distribution that characterizes the number of 0s generated by independent $\Ber(p)$ samples until the first 1. For $k=2$, we have $\eqref{eq:DLap-moments-2} = \frac{5}{2} \cdot \frac{\lambda^2-(1-\lambda)+1}{(1-\lambda)^2} <  \frac{5}{(1-\lambda)^2}$. For $k=4$, we have $\eqref{eq:DLap-moments-2} = \frac{5}{2} \cdot \frac{\lambda^4 +11\lambda^3 +11\lambda^2 +\lambda }{(1-\lambda)^4} < \frac{60}{(1-\lambda)^4}$.
\end{proof}

\subsection{Proofs for Preliminary Protocol}

\begin{thm}[Copy of Theorem \ref{thm:ut-prelim}]
Let $P=(R,\Sigma^{m,d}_{\eps,1/242},A)$ be the protocol specified by Algorithms \ref{alg:ut-randomizer} and \ref{alg:ut-analyzer}. Fix any $\eps \in (0,1)$. If $\lambda \gets \exp(-\eps)$, then $P$ is $(4\hat{\eps} + 2\eps)$-robustly differentially private. Moreover, there are choices of $\ell,m,\tau$ such that the sample complexity is $$N^* = O\left( \frac{d^{3/4}}{\alpha\eps} + \frac{d^{2/3}}{\alpha^{4/3}\eps^{2/3}} + \frac{d^{1/2}}{\alpha^2}\right)$$
\end{thm}

\begin{proof}
Privacy follows immediately from the privacy of our summation subroutine and composition. If $\vec{x},\vec{x}\,'$ are two neighboring datasets, there are exactly two values $j,\overline{j}$ such that $\hat{x}_{i,j} \neq \hat{x}'_{i,j}$ and $\hat{x}_{i,\overline{j}} \neq \hat{x}'_{i,\overline{j}}$. So, when running $P(\vec{x})$ and $P(\vec{x}\,')$ only two of the $d$ executions of the summation subroutine differ.

We now argue that the protocol correctly performs uniformity testing. To do so, let $q \gets 1/484d$, $g\gets 1$, and set the parameters $m,\tau$ according to Theorem \ref{thm:bounded-sum}. We will also enforce $\tau > \frac{\ln 10}{\ln (1/\lambda)} - 1$. We will derive the parameter $\ell$---the analyzer's threshold value--- and exact constants for the sample complexity $N^*$ later on.

We express the test statistic $Z'$ in terms of the true count of each $j$ in the dataset, which we denote $c_j(\vec{x}) := \sum_{i\in [n]} \ind{x_i=j}$. We will also use $\eta_j := \hat{c}_j - c_j(\vec{x})$ to denote the error in the count estimate.
\begin{align*}
    &Z' \\
    ={}& \frac{d}{N} \sum_{j\in[d]} \left( \hat{c}_j - \frac{N}{d} \right)^2 - \hat{c}_j \\
    ={}& \frac{d}{N} \sum_{j\in[d]} \left( c_j(\vec{x})+\eta_j - \frac{N}{d} \right)^2 - c_j(\vec{x}) - \eta_j \tag{By definition} \\
    ={}& \underbrace{\frac{d}{N} \sum_{j\in[d]} \left[ \left( c_j(\vec{x})- \frac{N}{d} \right)^2 - c_j(\vec{x}) \right]}_{Z} + \frac{d}{N} \sum_{j\in[d]} \eta^2_j + \frac{2d}{N} \sum_{j\in[d]} \eta_j\cdot \left( c_j(\vec{x}) - \frac{N}{d}\right) - \frac{d}{N} \sum_{j\in[d]} \eta_j \stepcounter{equation} \label{eq:ut-statistic} \tag{\theequation}
\end{align*}

\underline{Uniform Case}: We first prove that $Z' > \ell$ with probability $\leq t := 2/27$ when the underlying distribution is $\bU$, so that the output is ``not uniform'' with probability $\leq t$. In the other case, we'll show the corresponding probability is $\geq t' := 71/162$. Because $t'-t=\Omega(1)$, Remark \ref{rem:ut} applies.

The following analysis of $Z$ is immediate from work by Amin et al.\cite{AJM20} and Acharya, Daskalakis, and Kamath \cite{ADK15}:

\begin{clm}
\label{clm:ut-Z-uniform}
There exists a constant $\kappa$ such that if $N\geq \kappa \sqrt{d}/\alpha^2$, then in an execution of $P$ on $n\sim \Pois(N)$ samples from $\bU$,
\begin{align*}
    \ex{}{Z} &\leq \frac{\alpha^2 N}{500}\\
    \var{}{Z} &\leq \frac{\alpha^4 N^2}{500000}
\end{align*}
where $Z$ is defined in \eqref{eq:ut-statistic}.
\end{clm}

We now focus on the terms involving the privacy noise $\eta_j$. Let \textsf{notDLap} be the probability that, for some $j$, $\eta_j$ is not drawn from $\DLap_\tau(\lambda) * \DLap_\tau(\lambda)$. By a union bound and the construction of the analyzer, Claim \ref{clm:binary-sum-noise} implies that $\textsf{notDLap} \leq 1/242 + 2d \cdot \hat{q} = 1/242 + 1/242 = 1/81$. Conditioned on this event, there are independent random variables $\eta^{+}_j,\eta^{-}_j \sim \DLap_\tau(\lambda)$ such that
\begin{align*}
\eqref{eq:ut-statistic} ={}& Z + \frac{d}{N} \sum_{j\in[d]} \left( (\eta^{+}_j)^2 + (\eta^{-}_j)^2 \right) + \frac{2d}{N} \sum_{j\in[d]} \eta^{+}_j \eta^{-}_j \\
    &+ \underbrace{\frac{2d}{N} \sum_{j\in[d]} (\eta^{+}_j + \eta^{-}_j) \cdot \left( c_j(\vec{x}) - \frac{N}{d}\right)}_{T} - \left( \frac{d}{N} \sum_{j\in[d]} \eta^{+}_j + \eta^{-}_j \right) \stepcounter{equation} \label{eq:ut-statistic-T} \tag{\theequation}
\end{align*}

We now bound the variance of each term the above decomposition.
\begin{clm}
\label{clm:ut-DLap}
Fix any $\tau\in \N$ and $\lambda \in (0,1)$ such that $\tau > \frac{\ln 10}{\ln (1/\lambda)} - 1$. For any $d,N\in \N$, if we sample independent random variables $\eta^{+}_j,\eta^{-}_j \sim \DLap_\tau(\lambda)$ for all $j\in [d]$ then 
\begin{align*}
    \var{}{\frac{d}{N} \sum_{j\in[d]} \left( (\eta^{+}_j)^2 + (\eta^{-}_j)^2 \right)} &< \frac{120d^3}{N^2(1-\lambda)^4}\\
    \var{}{\frac{2d}{N} \sum_{j\in[d]} \eta^{+}_j \eta^{-}_j} &< \frac{100d^3}{N^2(1-\lambda)^4}\\
    \var{}{\frac{d}{N} \sum_{j\in[d]} \eta^{+}_j + \eta^{-}_j} &< \frac{10d^3}{N^2(1-\lambda)^2}
\end{align*}
\end{clm}

\begin{clm}
\label{clm:ut-cross-term}
In an execution of $P$ on $n\sim \Pois(N)$ samples from $\bU$,
\begin{align*}
    \ex{}{T} &= 0\\
    \var{}{T} &< \frac{40d^2}{N(1-\lambda)^2}
\end{align*}
where $T$ is defined in \eqref{eq:ut-statistic-T}.
\end{clm}

We prove these intermediary claims later. Using Chebyshev's inequality and a union bound, the following holds except with probability $\leq \textsf{notDLap} + 5\cdot (1/81) \leq 2/27 = t$:
\begin{align*}
\eqref{eq:ut-statistic-T} \leq{}& \ex{}{\eqref{eq:ut-statistic-T}} + 9\sqrt{\var{}{Z}} + 9\sqrt{ \var{}{\frac{d}{N} \sum_{j\in[d]} \left( (\eta^{+}_j)^2 + (\eta^{-}_j)^2 \right)} } + 9 \sqrt{\var{}{\frac{2d}{N} \sum_{j\in[d]} \eta^{+}_j \eta^{-}_j}} \\
    &+ 9 \sqrt{\var{}{T}} + 9 \sqrt{\var{}{\frac{d}{N} \sum_{j\in[d]} \eta^{+}_j + \eta^{-}_j}} \\
    <{}& \ex{}{\eqref{eq:ut-statistic-T}} + 9\cdot\left( \sqrt{\var{}{Z}} + \sqrt{\var{}{T}} + \frac{25d^{3/2}}{N(1-\lambda)^2} \right)\tag{Claim \ref{clm:ut-DLap}} \\
    <{}& \ex{}{\eqref{eq:ut-statistic-T}} + 9\cdot\left( \frac{\alpha^2 N}{\sqrt{500000}} + \frac{7d}{\sqrt{N}(1-\lambda)} + \frac{25d^{3/2}}{N(1-\lambda)^2} \right) \tag{Claims \ref{clm:ut-Z-uniform} and \ref{clm:ut-cross-term}} \\
    ={}& \frac{\alpha^2N}{500} + \frac{4d^2}{N}\ex{\eta\sim \DLap_\tau(\lambda)}{\eta^2} - \frac{2d^2}{N}\ex{\eta\sim \DLap_\tau(\lambda)}{\eta}\\
    &+9\cdot\left( \frac{\alpha^2 N}{\sqrt{500000}} + \frac{7d}{\sqrt{N}(1-\lambda)} + \frac{25d^{3/2}}{N(1-\lambda)^2} \right) \stepcounter{equation} \label{eq:ell} \tag{\theequation}
\end{align*}

\eqref{eq:ell} comes from linearity of expectation and the independence between $\eta^+_j,\eta^-_j$, along with Claims \ref{clm:ut-Z-uniform} and \ref{clm:ut-cross-term}. We assign the parameter $\ell$ to the right hand side of \eqref{eq:ell}.

\underline{Far-from-uniform Case}: We now show that, when the underlying distribution $\bD$ satisfies $\norm{\bD-\bU}_{\TV}>\alpha$, $Z' >\ell$ with probability at least $t'=71/162$. Observe that the gap $t'-t$ is the constant $59/162$, so the uniformity testing is solved.

The prior work by Amin et al. \cite{AJM20} and Acharya et al. \cite{ADK15} established the following facts about term $Z$ in this case:
\begin{clm}
\label{clm:ut-Z-far}
Fix any distribution $\bD$ where $\norm{\bD-\bU}_{\TV} > \alpha$. There exists a constant $\kappa$ such that if $N\geq \kappa \sqrt{d}/\alpha^2$, then in an execution of $P$ on $n\sim \Pois(N)$ samples from $\bD$,
\begin{align*}
    \ex{}{Z} &\geq \frac{\alpha^2 N}{5}\\
    \var{}{Z} &\leq \frac{\ex{}{Z}^2}{100}
\end{align*}
where $Z$ is defined in \eqref{eq:ut-statistic}.
\end{clm}

For the term $T$, note that $\eta_j=\eta^{+}_j+\eta^{-}_j$ is symmetrically distributed about zero. Invoking Claim 4.6 by Balcer et al., this implies $T$ is symmetrically distributed about zero so that $\pr{}{T \geq 0} \leq 1/2$.

We bound the other terms in \eqref{eq:ut-statistic-T}, we can again use Chebyshev's inequality. By a union bound, the following holds except with probability $\leq \textsf{notDLap} + 1/2 + 4\cdot (1/81) \leq 91/162 = (1-t')$:
\begin{align*}
\eqref{eq:ut-statistic-T} \geq{}& \ex{}{Z} + \frac{4d^2}{N}\ex{\eta\sim \DLap_\tau(\lambda)}{\eta^2} - \frac{2d^2}{N}\ex{\eta\sim \DLap_\tau(\lambda)}{\eta}\\
    &-9\sqrt{\var{}{Z}} - 9\sqrt{ \var{}{\frac{d}{N} \sum_{j\in[d]} \left( (\eta^{+}_j)^2 + (\eta^{-}_j)^2 \right)} } - 9 \sqrt{\var{}{\frac{2d}{N} \sum_{j\in[d]} \eta^{+}_j \eta^{-}_j}} \\
    &- 9 \sqrt{\var{}{\frac{d}{N} \sum_{j\in[d]} \eta^{+}_j + \eta^{-}_j}} \\
    >{}& \ex{}{Z} + \frac{4d^2}{N}\ex{\eta\sim \DLap_\tau(\lambda)}{\eta^2} - \frac{2d^2}{N}\ex{\eta\sim \DLap_\tau(\lambda)}{\eta}\\
    &-9\cdot\left(\sqrt{\var{}{Z}} +\frac{25d^{3/2}}{N(1-\lambda)^2} \right)\tag{Claim \ref{clm:ut-DLap}} \\
    \geq{}& \frac{\alpha^2 N}{50} + \frac{4d^2}{N}\ex{\eta\sim \DLap_\tau(\lambda)}{\eta^2} - \frac{2d^2}{N}\ex{\eta\sim \DLap_\tau(\lambda)}{\eta} -9\cdot\frac{25d^{3/2}}{N(1-\lambda)^2} \tag{Claim \ref{clm:ut-Z-far}}\\
    ={}& \ell + \frac{9\alpha^2 N}{500} - 9\cdot\left( \frac{\alpha^2 N}{\sqrt{500000}} + \frac{7d}{\sqrt{N}(1-\lambda)} \right) - 18\cdot \frac{25d^{3/2}}{N(1-\lambda)^2} \stepcounter{equation} \label{eq:ut-gap} \tag{\theequation}
\end{align*}
The last equality comes from substitution of our choice of $\ell$.

To complete the proof, it remains to argue that what is added to $\ell$ in \eqref{eq:ut-gap} is strictly positive.
\begin{align*}
& \frac{9\alpha^2 N}{500} - 9\cdot\left( \frac{\alpha^2 N}{\sqrt{500000}} + \frac{7d}{\sqrt{N}(1-\lambda)} \right) - 18\cdot \frac{25d^{3/2}}{N(1-\lambda)^2}\\
>{}& \frac{3\alpha^2 N}{100} - \frac{63d}{\sqrt{N}(1-\lambda)} - \frac{450d^{3/2}}{N(1-\lambda)^2} \\
\geq{}& \frac{3\alpha^2 N^*}{100} - \frac{63d}{\sqrt{N^*}(1-\lambda)} - \frac{450d^{3/2}}{N^*(1-\lambda)^2}  \tag{$N\geq N^*$} \\
={}& \frac{3\alpha^2 N^*}{100} - \frac{63de^\eps}{\sqrt{N^*}(e^\eps - 1)} - \frac{450d^{3/2}e^{2\eps} }{N^*(e^\eps - 1)^2} \tag{$\lambda \gets e^{-\eps}$} \\
={}& \underbrace{\left( \frac{3\alpha^2 N^*}{200} - \frac{63de^\eps}{\sqrt{N^*}(e^\eps - 1)} \right)}_{A} + \underbrace{\left( \frac{3\alpha^2 N^*}{200} - \frac{450d^{3/2}e^{2\eps} }{N^*(e^\eps - 1)^2} \right)}_{B}
\stepcounter{equation} \label{eq:ut-N-star} \tag{\theequation}
\end{align*}
We finally set $N^*$ to be equal to $\frac{\kappa \sqrt{d}}{\alpha^2} + \frac{174d^{3/4}e^\eps}{\alpha (e^\eps-1)} + \frac{261d^{2/3}e^{2\eps/3}}{\alpha^{4/3}(e^\eps-1)^{2/3}}$ where $\kappa$ is the constant in Claims \ref{clm:ut-Z-uniform} and \ref{clm:ut-Z-far}. Term $A$ is strictly positive because $N^* > \frac{261d^{2/3}e^{2\eps/3}}{\alpha^{4/3}(e^\eps-1)^{2/3}}$ and Term $B$ is strictly positive because $N^* > \frac{174d^{3/4}e^\eps}{\alpha (e^\eps-1)}$
\end{proof}

\begin{proof}[Proof of Claim \ref{clm:ut-DLap}]
The bounds are immediate from independence and the bounds we derived on the moments of the truncated discrete Laplace distribution:
\begin{align*}
    \var{}{\frac{d}{N} \sum_{j\in[d]} \left( (\eta^{+}_j)^2 + (\eta^{-}_j)^2 \right)} &= \frac{2d^3}{N^2} \cdot \var{\eta \sim \DLap_\tau(\lambda)}{\eta^2} \tag{Independence} \\
    &= \frac{2d^3}{N^2} \cdot \left( \ex{\eta \sim \DLap_\tau(\lambda)}{\eta^4} - \ex{\eta \sim \DLap_\tau(\lambda)}{\eta^2}^2 \right)\\
    &\leq \frac{2d^3}{N^2} \cdot \ex{\eta \sim \DLap_\tau(\lambda)}{\eta^4} \\
    &\leq \frac{120d^3}{N^2(1-\lambda)^4} \tag{Claim \ref{clm:DLap}}
\end{align*}

\begin{align*}
\var{}{\frac{2d}{N} \sum_{j\in[d]} \eta^{+}_j \eta^{-}_j} &= \frac{4d^3}{N^2} \cdot \var{\eta \sim \DLap_\tau(\lambda)}{\eta}^2 \tag{Independence}\\
    &= \frac{4d^3}{N^2} \cdot \left( \ex{\eta \sim \DLap_\tau(\lambda)}{\eta^2} - \ex{\eta \sim \DLap_\tau(\lambda)}{\eta}^2 \right)^2\\
    &< \frac{100d^3}{N^2(1-\lambda)^4} \tag{Claim \ref{clm:DLap}}
\end{align*}

\begin{align*}
\var{}{\frac{d}{N} \sum_{j\in[d]} \eta^{+}_j + \eta^{-}_j} &= \frac{2d^3}{N^2} \cdot \var{\eta \sim \DLap_\tau(\lambda)}{\eta} \tag{Independence}
    &= \frac{2d^3}{N^2} \cdot \left( \ex{\eta \sim \DLap_\tau(\lambda)}{\eta^2} - \ex{\eta \sim \DLap_\tau(\lambda)}{\eta}^2 \right) \\
    &< \frac{10d^3}{N^2(1-\lambda)^2}
\end{align*}
\end{proof}

\begin{clm*}[Copy of Claim \ref{clm:ut-cross-term}]
In an execution of $P$ on $n\sim \Pois(N)$ samples from $\bU$,
\begin{align*}
    \ex{}{T} &= 0\\
    \var{}{T} &< \frac{40d^2}{N(1-\lambda)^2}
\end{align*}
where $T$ is defined in \eqref{eq:ut-statistic-T}.
\end{clm*}
\begin{proof}
By linearity of expectation,
\begin{align*}
\ex{}{T} &= \frac{2d}{N} \sum_{j \in [d]} \ex{}{(\eta^+_j + \eta^-_j) \cdot \left( c_j(\vec{x}) - \frac{N}{d} \right)} \\
    &= \frac{2d}{N} \sum_{j \in [d]} \ex{}{(\eta^+_j + \eta^-_j) } \cdot \ex{}{\left( c_j(\vec{x}) - \frac{N}{d} \right)} \tag{Independence} \\
    &= \frac{4d}{N} \sum_{j \in [d]} \ex{\eta \sim \DLap_\tau(\lambda)}{\eta} \cdot \ex{}{\left( c_j(\vec{x}) - \frac{N}{d} \right)} \tag{Linearity}\\
    &= 0 \tag{Claim \ref{clm:DLap}}
\end{align*}

To bound the variance of $T$, we will make use of the following fact:
\begin{fact}
\label{fact:poissonization}
If $n\sim \Pois(N)$ and $\vec{x}\sim \bU^n_{[d]}$ then for all $j$, $c_j(\vec{x})$ is an independent sample from $\Pois(N/d)$.
\end{fact}

Because noise $\eta^{+}_j+\eta^{-}_j$ is independent of the data $\vec{x}$,
\begin{align*}
\var{}{T} &= \frac{4d^2}{N^2} \sum_{j \in [d]} \var{}{(\eta^+_j + \eta^-_j)} \cdot \var{}{\left( c_j(\vec{x}) - \frac{N}{d} \right)} \\
    &= \frac{8d^2}{N^2}  \sum_{j\in [d]} \var{\eta \sim \DLap_\tau(\lambda)}{\eta} \cdot \var{}{\left( c_j(\vec{x}) - \frac{N}{d} \right)} \tag{$\eta^+_j$ independent of $\eta^-_j$}\\
    &= \frac{8d^2}{N} \cdot \var{\eta \sim \DLap_\tau(\lambda)}{\eta} \tag{Fact \ref{fact:poissonization}}\\
    &< \frac{40d^2}{N(1-\lambda)^2} \tag{Claim \ref{clm:DLap}}
\end{align*}

This concludes the proof.
\end{proof}

\subsection{Proofs for Final Protocol}
Many of the steps to prove Theorem \ref{thm:ut-final} are verbatim from prior work, but we reproduce them here for completeness.

\begin{thm}[Copy of Theorem \ref{thm:ut-final}]
Let $P=(R,\Sigma^{m,\hat{d}}_{\hat{\eps},1/200},A)$ be the protocol specified by Algorithms \ref{alg:ut-final-randomizer} and \ref{alg:ut-final-analyzer}. Fix any $\eps \in (0,1)$. If $\lambda \gets \exp(-\eps)$, then $P$ is $(4\hat{\eps} + 2\eps)$-robustly differentially private. Moreover, there are choices of $\hat{d}, \ell, m, \tau$ such that the sample complexity is $$N^* = O\left( \frac{d^{2/3}}{\alpha^{4/3}\eps^{2/3}} + \frac{d^{1/2}}{\alpha\eps} + \frac{d^{1/2}}{\alpha^2} \right)$$
\end{thm}
\begin{proof}
Because the protocol simply executes a private protocol on randomly binned data, privacy is immediately inherited.

We assign $\hat{d}$ according to the following rule:
$$
\hat{d} ~~ \begin{cases}
2 & \textrm{if } \frac{d^{2/3}\eps^{4/3}}{\alpha^{4/3}} < 2 \\
d & \textrm{if } \frac{d^{2/3}\eps^{4/3}}{\alpha^{4/3}} > d \\
\frac{d^{2/3}\eps^{4/3}}{\alpha^{4/3}} & \textrm{otherwise}\\
\end{cases}
$$

Let $\hat{\alpha}$ be the value $\alpha \cdot \frac{\sqrt{\hat{d}}}{477 \sqrt{10 d}}$. We set the parameters $\ell, m, \tau$ so that the invocation of the preliminary protocol (pseudocode in Algorithms \ref{alg:ut-randomizer} and \ref{alg:ut-analyzer}) solves $\hat{\alpha}$-uniformity testing with sample complexity $\hat{N}^*$ (for the compressed universe $[\hat{d}]$, not $[d]$).

\underline{Sample Complexity}: From Theorem \ref{thm:ut-prelim}, the sample complexity $N^*$ for $\hat{\alpha}$-uniformity testing the universe $[\hat{d}]$ is 

\begin{align*}
N^* &= O\left(  \frac{\hat{d}^{3/4}}{\hat{\alpha}\eps} + \frac{\hat{d}^{2/3}}{\hat{\alpha}^{4/3}\eps^{2/3}} + \frac{\hat{d}^{1/2}}{\hat{\alpha}^2} \right) \\
    &= O\left( \underbrace{\frac{d^{1/2}\hat{d}^{1/4}}{\alpha\eps}}_{T_1} + \underbrace{\frac{d^{2/3}}{\alpha^{4/3}\eps^{2/3}}}_{T_2} + \underbrace{\frac{d}{\alpha^2 \hat{d}^{1/2}}}_{T_3} \right)
\end{align*}

To arrive at our desired bound, we split into cases of $\hat{d}$.

\textit{Case 1:} $\hat{d} = 2$, which means $\frac{d^{2/3}\eps^{4/3}}{\alpha^{4/3}} < 2$. Rearranging terms, this means $d^{1/2} = O(\alpha/\eps)$ and $d^{1/6} = O(\alpha^{1/3}/\eps^{1/3})$. Thus,
\begin{align*}
	T_1 + T_2 + T_3 =&\ O\left(\frac{d^{1/2}}{\alpha \eps} + \frac{d^{1/2}\cdot d^{1/6}}{\alpha^{4/3}\eps^{2/3}} + \frac{d^{1/2} \cdot d^{1/2}}{\alpha^2} \right) \\
	=&\ O\left(\frac{d^{1/2}}{\alpha \eps} + \frac{d^{1/2}}{\alpha \eps} + \frac{d^{1/2}}{\alpha \eps}\right) \\
	=&\ O\left(\frac{d^{1/2}}{\alpha\eps}\right).
\end{align*}

\textit{Case 2}: $\hat d = d$, so $d < \tfrac{d^{2/3}\eps^{4/3}}{\alpha^{4/3}}$. Rearranging terms, $d^{3/4} < \tfrac{d^{1/2}\eps}{\alpha}$ and $d^{1/6} < \tfrac{\eps^{2/3}}{\alpha^{2/3}}$. Thus,
\begin{align*}
T_1 + T_2 + T_3 =&\ O\left(\frac{d^{3/4}}{\alpha \eps} + \frac{d^{2/3}}{\alpha^{4/3}\eps^{2/3}} + \frac{d^{1/2}}{\alpha^2} \right) \\
	=&\ O\left(\frac{d^{1/2}}{\alpha^2} + \frac{d^{2/3}}{\alpha^{4/3}\eps^{2/3}} \right) \\
	=&\ O\left(\frac{d^{1/2}}{\alpha^2} + \frac{d^{1/2} \cdot d ^{1/6}}{\alpha^{4/3}\eps^{2/3}} \right) \\
	=&\ O\left(\frac{d^{1/2}}{\alpha^2}\right).
\end{align*}

\textit{Case 3}: $\hat{d} = \tfrac{d^{2/3} \eps^{4/3}}{\alpha^{4/3}}$. By substitution,
\begin{align*}
T_1 + T_2 + T_3 =&\ O\left(\frac{d^{1/2} (d^{2/3} \eps^{4/3} \alpha^{-4/3})^{1/4} }{\alpha\eps } + \frac{d^{2/3}}{\alpha^{4/3}\eps^{2/3}} + \frac{d}{\alpha^2 (d^{2/3} \eps^{4/3} \alpha^{-4/3})^{1/2}}\right) \\
	=&\ O\left(\frac{d^{2/3} }{\alpha^{4/3}\eps^{2/3} } \right).
\end{align*}
	
\underline{Correctness}: By construction, we are feeding samples from $\bD$ into $\texttt{Coarsen}(\cdot, G)$ and then passing the result into the preliminary protocol. This means the sampling distribution is transformed from $\bD$ into $\bD_G$. By a slight generalization of Remark \ref{rem:ut}, we can conclude the following for $n\sim \Pois(O(N^*))$ samples from $\bD_G$.
\begin{itemize}
    \item the preliminary protocol reports ``not uniform'' with probability at most $1/9540$ when $\bD_G = \bU_{[\hat{d}]}$ 
    \item the preliminary protocol reports ``not uniform'' with probability at least $9539/9540$ when $\norm{\bD_G - \bU_{[\hat{d}]}}_\TV > \hat{\alpha}$ 
\end{itemize}

If $\bD=\bU_{[d]}$ and $\hat{d}$ is a factor of $d$, then observe that $\bD_G$ is equal to $\bU_{[\hat{d}]}$ for any choice of $G$. This follows from a simple computation: $$\pr{}{\bD_G = \hat{j}} = \sum_{j \in G_{\hat{j}}} \pr{}{\bD = j} = \sum_{j \in G_{\hat{j}}} 1/d = 1/\hat{d}.$$ So we can immediately invoke Theorem \ref{thm:ut-prelim} and conclude that the probability that the final protocol reports ``not uniform'' is at most $1/9540$. If $\hat{d}$ is \emph{not} a factor of $d$, we can tweak the protocol in the following way: mix user data with $\bU_{\tilde{d}}$ where $\tilde{d}=O(d)$ such that $\hat{d}$ is a factor of $\tilde{d}$.

If $\norm{\bD- \bU_{[d]} }_\TV>\alpha$ then by Lemma~\ref{lem:compression}, $\norm{\bD_G - \bU_{[\hat{d}]}}_\TV > \hat{\alpha}$ except with probability $\leq 953/954$ over the randomness of $G$. By a union bound, the final protocol returns ``non-uniform'' except with probability $\leq 9531/9540$. 

Thus, the gap between the failure probability in the uniform case and the success probability in the far-from-uniform case is $9/9540 - 1/9540$, a constant. This concludes the proof.
\end{proof}

\end{document}